\documentclass{jfp}

\usepackage{amsmath}
\usepackage{amssymb}
\usepackage{stmaryrd}
\usepackage[all]{xy}
\usepackage{subfigure}
\usepackage{url}
\usepackage{util}

\allowdisplaybreaks[2]

\newenvironment{smath}{\[}{\]}

\newenvironment{prog}{\begin{array}[t]{@{}l@{}}}{\end{array}}

\newcommand{\hbra}{
\hbox to .995 \textwidth{\vrule width0.3mm height 1.8mm depth-0.3mm
                    \leaders\hrule height1.8mm depth-1.5mm\hfill
                    \vrule width0.3mm height 1.8mm depth-0.3mm}}
\newcommand{\hket}{
\hbox to .995 \textwidth{\vrule width0.3mm height1.5mm
                    \leaders\hrule height0.3mm\hfill
                    \vrule width0.3mm height1.5mm}}

\newcommand{\ratio}{.4}
\newenvironment{display}[1]{\begin{tabbing}
  \hspace{1.5em} \= \hspace{\ratio\linewidth-1.5em} \= \hspace{1.5em} \= \kill
  {\bfseries#1}\\[-.8ex]
  \hbra\\[-.8ex]
  }{\\[-.8ex]\hket
  \end{tabbing}}
\newcommand{\entry}[2]{\>$#1$\>\>#2}
\newcommand{\clause}[2]{$#1$\>\>#2}
\newcommand{\category}[2]{\clause{#1::=}{#2}}

\newcommand{\toappend}{}

\newenvironment{append}
  {\let\OldtheTHEOREM\theTHEOREM
   \begin{renewcommand}{\theTHEOREM}{\OldtheTHEOREM{}\toappend}}
  {\end{renewcommand}}

\newtheorem{THEOREM}{Theorem}[section]
\newtheorem{LEMMA}[THEOREM]{Lemma}
\newtheorem{COROLLARY}[THEOREM]{Corollary}
\newtheorem{PROPOSITION}[THEOREM]{Proposition}
\newtheorem{DEFINITION}[THEOREM]{Definition}

\newtheorem{EXAMPLE}[THEOREM]{Example}
\newtheorem{REMARK}[THEOREM]{Remark}

\newenvironment{theorem}
 {\begin{append}\begin{THEOREM}}
 {\end{THEOREM}\end{append}}
\newenvironment{lemma}
 {\begin{append}\begin{LEMMA}}
 {\end{LEMMA}\end{append}}

\newenvironment{proposition}
 {\begin{append}\begin{PROPOSITION}}
 {\end{PROPOSITION}\end{append}}

\newenvironment{example*}
  {\begin{append}\begin{EXAMPLE} \rm}
  {\end{EXAMPLE}\end{append}}
\newenvironment{remark*}
  {\begin{append}\begin{REMARK} \rm}
  {\end{REMARK}\end{append}}

\newenvironment{oldtheorem}[1]
  {\begin{renewcommand}{\theTHEOREM}{\ref{#1}\toappend}}
  {\end{renewcommand}\addtocounter{THEOREM}{-1}}

\def\squareforqed{\hbox{\rlap{$\sqcap$}$\sqcup$}}
\def\wbox{\ifmmode\squareforqed\else{\unskip\nobreak\hfil
\penalty50\hskip1em\null\nobreak\hfil\squareforqed
\parfillskip=0pt\finalhyphendemerits=0\endgraf}\fi}

\newcommand{\expr}[1]{\texttt{#1}}
\newcommand{\itype}[1]{\textit{#1}}
\newcommand{\type}[1]{\texttt{#1}}

\newcommand{\abst}[2][]{\langle #2 \rangle_{A_{#1}}}
\newcommand{\conc}[2][]{\langle #2 \rangle_{C_{#1}}}
\newcommand{\namesigma}[1]{\langle #1 \rangle_{N}}
\newcommand{\auxenc}[1]{\langle #1 \rangle_{X}}
\newcommand{\embed}[1]{\mathit{inj}(#1)}
\newcommand{\new}{{\scriptscriptstyle\rm new}}
\newcommand{\parent}{{\scriptscriptstyle\rm parent}}

\newcommand{\subt}{<:}
\newcommand{\subto}{\mathord{<:}}
\newcommand{\Fsub}{F_{\subto}}

\newcommand{\scalc}{\lambda^{\scriptscriptstyle\rm DM}_{\scriptscriptstyle\subto}}
\newcommand{\sj}{\vdash_{\scriptscriptstyle\subto}}
\newcommand{\sjudget}[3]{#1\sj #2:#3}
\newcommand{\sjudgest}[2]{\sj #1\subt #2}
\newcommand{\slra}{\longrightarrow_{\scriptscriptstyle\subto}}
\newcommand{\tcalc}{\lambda^{\scriptscriptstyle\rm DM}_{\scriptscriptstyle\sf T}}
\newcommand{\tj}{\vdash_{\scriptscriptstyle\sf T}}
\newcommand{\tjudget}[3]{#1\tj #2:#3}
\newcommand{\tlra}{\longrightarrow_{\scriptscriptstyle\sf T}}

\newcommand{\ForT}[2]{\forall #1(#2)}
\newcommand{\For}[3]{\forall #1\subto #2(#3)}
\newcommand{\Forn}[3]{\forall #1_1\subto #2_1,\dots,#1_n\subto #2_n(#3)}
\newcommand{\Lamn}[3]{\Lambda #1_1\subto #2_1,\dots,#1_n\subto #2_n(#3)}
\newcommand{\LamT}[2]{\Lambda #1(#2)}
\newcommand{\LamTn}[2]{\Lambda #1_1\dots,#1_n(#2)}
\newcommand{\lam}[3]{\lambda #1\mathord{:}#2(#3)}
\newcommand{\Let}[3]{\mbox{\textbf{let}}~#1=#2~\mbox{\textbf{in}}~#3}
\newcommand{\Lam}[3]{\Lambda #1\subto #2(#3)}

\newcommand{\RuleSide}[3]{
{\Rule{#1}{#2}}{~~#3}
}
\newcommand{\Rule}[2]{
  \begin{array}{c}
  #1 \\\hline
  #2
  \end{array}}

\newcommand{\GAP}{2ex}

\newcommand{\intension}[1]{\llbracket#1\rrbracket}
\newcommand{\cA}{\mathcal{A}}
\newcommand{\cB}{\mathcal{B}}
\newcommand{\cC}{\mathcal{C}}
\newcommand{\cE}{\mathcal{E}}
\newcommand{\cT}{\mathcal{T}}

\newcommand{\leqo}{\subto}
\renewcommand{\emptyset}{\varnothing}
\newcommand{\emptyseq}{\langle\,\rangle}

\newcommand{\tyalpha}{\texttt{'a}}
\newcommand{\tybeta}{\texttt{'b}}

\newcommand{\tytimes}{\mathrel{\texttt{*}}}
\newcommand{\tyto}{\mathrel{\texttt{->}}}

\newcommand{\teq}{\triangleq}

 \title[Phantom Types and Subtyping]
    {Phantom Types and Subtyping}
 \author[M. Fluet and R. Pucella]
        {Matthew Fluet\\
         Cornell University \\
         Ithaca, NY 14853 USA\\
         \email{fluet@cs.cornell.edu} 
   \and
         Riccardo Pucella\\
         Northeastern University\\
         Boston, MA 02115 USA\\
         \email{riccardo@ccs.neu.edu}}
        
 \date{}

\begin{document}
\maketitle

\begin{abstract}
  We investigate a technique from the literature, called the
  phantom-types technique, that uses parametric polymorphism, type
  constraints, and unification of polymorphic types to model a
  subtyping hierarchy.  Hindley-Milner type sys\-tems, such as the one
  found in Standard ML, can be used to enforce the subtyping relation, at least for first-order values.  We show
  that this technique can be used to encode any finite subtyping
  hierarchy (including hierarchies arising from multiple interface
  inheritance).  We formally demonstrate the suitability of the
  phantom-types technique for capturing first-order subtyping by exhibiting a
  type-preserving translation from a simple calculus with bounded
  polymorphism to a calculus embodying the type system of SML.
\end{abstract}

\section{Introduction}
\label{s:introduction}

It is well known that traditional type systems, such as the one found
in Standard ML~\cite{Milner97}, with parametric polymorphism and type
constructors, can be used to capture program properties beyond those
naturally associated with a Hindley-Milner type
system~\cite{Milner78}.  For concreteness, let us review a simple
example, due to Leijen and Meijer~\shortcite{Leijen99}.  Consider a
type of atoms, either booleans or integers, that can be easily
represented with an algebraic datatype:
\begin{center}
\begin{minipage}[t]{0.750\textwidth}
\enablettchars
datatype atom = I of int | B of bool.
\end{minipage}
\end{center}
There are a number of operations that we may perform on such atoms
(see Figure~\ref{f:unsafeops}).  When the domain of an operation is
restricted to only one kind of atom, as with \expr{conj} and
\expr{double}, a run-time check must be made and an error or exception
reported if the check fails.

\newsavebox{\unsafeopsBox}
\begin{lrbox}{\unsafeopsBox}
\begin{minipage}[t]{0.48\textwidth}
\scriptsize\enablettchars
datatype atom = I of int | B of bool
~
~
fun mkInt (i:int):atom = I (i)
fun mkBool (b:bool):atom = B (b)

fun toString (v:atom):string =
  (case v
    of I (i) => Int.toString (i)
     | B (b) => Bool.toString (b))
fun double (v:atom):atom =
  (case v
    of I (i) => I (i * 2)
     | \_ => raise Fail "type mismatch")
fun conj (v1:atom,
          v2:atom):atom =
  (case (v1,v2)
    of (B (b1), B (b2)) => B (b1 andalso b2)
     | \_ => raise Fail "type mismatch")
~
\end{minipage}
\end{lrbox}

\newsavebox{\safeopsBox}
\begin{lrbox}{\safeopsBox}
\begin{minipage}[t]{0.48\textwidth}
\scriptsize\enablettchars
datatype atom = I of int | B of bool
datatype 'a safe\_atom = W of atom
~
fun mkInt (i:int):int safe\_atom = W (I (i))
fun mkBool (b:bool):bool safe\_atom = W (B (b))

fun toString (v:'a safe\_atom):string = 
  (case v
    of W (I (i)) => Int.toString (i)
     | W (B (b)) => Bool.toString (b))
fun double (v:int safe\_atom):int safe\_atom =
  (case v
    of W (I (i)) => W (I (i * 2))
     | \_ => raise Fail "type mismatch")
fun conj (v1:bool atom,
          v2:bool atom):bool atom =
  (case (v1,v2)
    of (W (B (b1)), W (B (b2))) => 
               W (B (b1 andalso b2))
     | \_ => raise Fail "type mismatch")
~
\end{minipage}
\end{lrbox}

\begin{figure}
\hrule
\center{
\subfigure[\label{f:unsafeops}Unsafe operations]{%
\usebox{\unsafeopsBox}
}\hspace{\fill}%
\subfigure[\label{f:safeops}Safe operations]{%
\usebox{\safeopsBox}
}}%
\hrule
\caption{Atom operations}
\end{figure}

One aim of static type checking is to reduce the number of run-time
checks by catching type errors at compile time.  Of course, in the
example above, the SML type system does not consider \expr{conj (mkInt
  3, mkBool true)} to be ill-typed; evaluating this expression will
simply raise a run-time exception.

If we were working in a language with subtyping, we would like to
consider integer atoms and boolean atoms as distinct subtypes of the
general type of atoms and use these subtypes to refine the types of
the operations.  
Then the type system would report, at compile time, a type error in the expression
\expr{double (mkBool false)}.
Fortunately, we can write the operations in a way that utilizes the
SML type system to do just this.
We introduce a new datatype that represents ``safe'' atoms, which 
is a simple wrapper around the datatype for atoms:
\begin{center}
\begin{minipage}[t]{0.750\textwidth}
\enablettchars
datatype $\tyalpha$ safe\_atom = W of atom
\end{minipage}
\end{center}
and constrain the types of the operations (see
Figure~\ref{f:safeops}).  We use the superfluous type variable $\tyalpha$ in the
datatype definition to encode information about the kind of atom.
(Because instantiations of this type variable do not contribute to the
run-time representation of atoms, it is called a \emph{phantom type}.)
The type \type{int safe\_atom} is used to represent integer atoms and
\type{bool safe\_atom} is used to represent boolean atoms.  Now, the
expression \expr{conj (mkInt 3, mkBool true)} results in a compile-time
type error, because the types \type{int safe\_atom} and \type{bool safe\_atom} do
not unify.  (Observe that our use of \type{int} and \type{bool} as
phantom types is arbitrary; we could have used any two types that do
not unify to make the integer versus boolean distinction.)  On the
other hand, both \expr{toString (mkInt 3)} and \expr{toString (mkBool
 true)} are well-typed; the \expr{toString} operation can be applied to any atom.
Note that we had to wrap the \type{atom} type in another datatype;
the next section will explain why.\footnote{We could have simply
defined \type{safe\_atom} as:
\begin{center}
\begin{minipage}[t]{0.750\textwidth}
\enablettchars
datatype $\tyalpha$ safe\_atom = I of int | B of bool
\end{minipage}
\end{center}
but for the sake of uniformity with the techniques presented in the
next section, we use the slightly more verbose wrapping using a
\expr{W} constructor.}

The example above used a datatype as the representation of values
manipulated by ``unsafe'' operations,
and a wrapped version of the datatype to enforce
safety. 
However, the underlying representation
need not be a datatype. Consider a common instance of the
problem, where 
we wish
is to manipulate operating-system values, such as
sockets. These are typically accessed via a foreign-function 
interface and they are typically represented by a 32-bit integer
value (either representing a pointer or a handle in a table kept by
the operating system). A number of primitive operations are provided
through the foreign-function interface for handling those sockets (see
Figure~\ref{f:sockets-unsafe}).  However, while
the SML representation of a socket is just a 32-bit integer,
the operating system often distinguishes internally between different
kinds of sockets, for instance, between UDP sockets and TCP sockets,
and operations specific to UDP sockets cause run-time exceptions (at
the operating-system level) when supplied with a TCP socket. For
instance, the operation \expr{sendUDP} expects a UDP socket and a
string to send on the socket. This is exactly the kind of check that
occurs in the atom example above, except it is performed automatically
by the operating system rather than the code. Other operations, such
as \expr{close}, work with all sockets, and therefore, there is an
implicit subtyping relation among sockets, UDP sockets, and TCP
sockets. We can enforce the appropriate use of sockets statically by
defining new types:
\begin{center}
\begin{minipage}[t]{0.750\textwidth}
\enablettchars
datatype udp = UDP
datatype tcp = TCP
datatype $\tyalpha$ safe\_sock = W of Word32.word
\end{minipage}
\end{center}
and constraining the types of the operations appropriately (see
Figure~\ref{f:sockets-safe}). 
Note that we again use
a superfluous type variable in the definition of the type
\type{safe\_sock} to allow us to constrain the type of the
operations. We can now supply appropriate types to versions of safe
operations on sockets. (Note once again that we had to wrap the
\type{Word32.word} type in a datatype.)

\newsavebox{\unsafeopssockBox}
\begin{lrbox}{\unsafeopssockBox}
\begin{minipage}[t]{0.450\textwidth}
\scriptsize\enablettchars
type sock = Word32.word
~
~
~
fun makeUDP (addr:string):sock = 
  ffiMakeUDP (addr)
fun makeTCP (addr:string):sock = 
  ffiMakeTCP (addr)
~
fun sendUDP (s:sock,text:string):unit = 
  ffiSendUDP (s,text)
fun sendTCP (s:sock,text:string):unit = 
  ffiSendTCP (s,text)
~
fun close (s:sock):unit = 
  ffiClose (s)
~
\end{minipage}
\end{lrbox}

\newsavebox{\safeopssockBox}
\begin{lrbox}{\safeopssockBox}
\begin{minipage}[t]{0.500\textwidth}
\scriptsize\enablettchars
datatype udp = UDP
datatype tcp = TCP
datatype 'a safe\_sock = W of Word32.word
~
fun makeUDP (addr:string):udp safe\_sock = 
  W (ffiMakeUDP (addr))
fun makeTCP (addr:string):tcp safe\_sock = 
  W (ffiMakeTCP (addr))
~
fun sendUDP (s:udp safe\_sock,text:string):unit = 
  (case s of W (w) => ffiSendUDP (w,text))
fun sendTCP (s:tcp safe\_sock,text:string):unit = 
  (case s of W (w) => ffiSendTCP (w,text))
~
fun close (s:'a safe\_sock):unit = 
  (case s of W (w) => ffiClose (w))
~
\end{minipage}
\end{lrbox}

\begin{figure}
\hrule
\center{
\subfigure[\label{f:sockets-unsafe}Unsafe operations]{%
\usebox{\unsafeopssockBox}
}\hspace{\fill}%
\subfigure[\label{f:sockets-safe}Safe operations]{%
\usebox{\safeopssockBox}
}}%
\hrule
\label{f:sockets}
\caption{Socket operations}
\end{figure}

This is the essence of the technique explored in this paper: using a
free type variable to encode subtyping information for first-order
values, and using an SML-like type system to enforce the subtyping on
those values. (We focus on first-order subtyping in this paper;
Section~\ref{s:bounded} explains why.) 
This ``phantom types'' technique, where user-defined restrictions are
reflected in the constrained types of values and functions, underlies
many interesting uses of type systems.
It has been used to derive early implementations of extensible
records~\cite{Wand87,Remy89,Burton90}, to provide a safe and flexible
interface to the Network Socket API~\cite{Reppy96}, to interface to
COM components~\cite{Finne99}, to type embedded compiler
expressions~\cite{Leijen99,Elliott00}, to record sets of effects in
type-and-effect type systems~\cite{Pessaux99}, to embed a
representation of the C type system in SML~\cite{Blume01},
and to encode data structure invariants~\cite{FluetPucella05}.

This paper makes 
a number of contributions
to the extant literature on phantom types. The first contribution is
to describe a general procedure for applying the phantom-types
technique to subtyping, generalizing all the known uses of phantom
types of which we are aware.  This procedure relies on an appropriate
encoding of the subtyping hierarchy. We study different classes of
encodings for different kinds of hierarchies. 
Next, we formalize this use of phantom types and prove its
correctness.  We present a type-preserving translation from a calculus
with subtyping to a calculus with let-bounded polymorphism, using the
procedure described earlier.  The kind of subtyping that can be
captured turns out to be an interesting variant of bounded
polymorphism~\cite{Cardelli94}, with a very restricted subsumption
rule.  

This paper is structured as follows. In the next section, we describe
a simple recipe for deriving an interface enforcing a given subtyping
hierarchy. The interface is parameterized by an encoding, via phantom
types, of the subtyping hierarchy.  In Section~\ref{s:encodings}, we
examine different encodings for hierarchies. We also address the issue
of extensibility of the encodings.  In Section~\ref{s:bounded}, we
extend the recipe to capture a limited form of bounded polymorphism.
In Section~\ref{s:formal}, we formally define the kind of subtyping
captured by our encodings by giving a simple calculus with subtyping
and showing that our encodings provide a type-preserving translation
to a variant of the Damas-Milner calculus, embodying the essence of
the SML type system.  
We conclude with some problems inherent to the approach and a
consideration of future work.
The formal details of the calculi we introduce in
Section~\ref{s:formal} as well as the proofs of our results can be
found in the appendices.  

\section{From Subtyping to Polymorphism}
\label{s:subtyping}

The examples in the introduction has the following features: an
underlying \emph{base type} of values 
(the original type \type{atom}
and the type \type{Word32.word} for sockets),  
a set of
primitive
operations on values of the base type, and \emph{sorts} of this base
type that correspond to the sensible domains of the operations. The
sorts of the base type form a hierarchy capturing the subtyping
inherent in the sorts.  The subtyping hierarchy corresponding to the
atom
example is as follows:
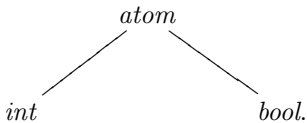
\begin{displaymath}
\begin{xy}
\xymatrix{
{} & \mbox{\itype{atom}} \ar@{-}[dl] \ar@{-}[dr] & \\
\mbox{\itype{int}} & & \mbox{\itype{bool}.}
}
\end{xy}
\end{displaymath}
(We assume there is a sort corresponding to the base type as a whole,
always the top of the hierarchy, capturing the intuition that every
sort is a subtype of the base type.)  The subtyping hierarchy is
modeled by assigning a type to every sort in the hierarchy.  For
instance, integer atoms with sort \itype{int} are encoded by the SML
type \type{int safe\_atom}.  The appropriate use of polymorphic type
variables in the type of an operation indicates the maximal type in
the domain of the operation.  For instance, the operation
\expr{toString} has the conceptual type \type{{\normalfont\itype{atom}} $\tyto$ string}
which is encoded by the SML type \type{$\tyalpha$ safe\_atom $\tyto$
string}.  The key observation is the use of type unification to
enforce the subtyping hierarchy: an \type{int safe\_atom} can be
passed to a function expecting an \type{$\tyalpha$ safe\_atom}, because
these types unify.

In this section, we show by means of an example, given a base
type $\tau_b$, a set of sorts of $\tau_b$ (forming a hierarchy), and
operations expressed in terms of the sorts of $\tau_b$, how to derive:
\begin{itemize}
\item a safe SML signature which uses phantom types to encode the
subtyping between sorts, and
\item a safe implementation from the unsafe implementation.   
\end{itemize}
By safety here, we mean  that the interface guarantees that no
primitive operation is ever supplied a value outside its domain; we
return to this point in Section~\ref{s:safe}, and make this guarantee
precise in Section~\ref{s:formal}. 

All values share the same underlying representation (the base type
$\tau_b$) and each operation has a single implementation that acts on
this underlying representation.  The imposed subtyping captures
restrictions that arise because of some external knowledge about the
semantics of the operations; intuitively, it captures a ``real''
subtyping relationship that is not exposed by 
the representation of 
the abstract type.

We must emphasize at this point that we are concerned only with
subtyping of values passed to primitive operations, where the values
are base values as opposed to higher-order values such as
functions. Therefore, we are interesting in first-order subtyping
only. Of course, since SML is higher-order, we must say something
about the subtyping on higher-order values induced by the subtyping on
the base values. The subtyping relation on higher-order values will
turn out to be severely restricted. We return to this point in
Section~\ref{s:bounded}. 

\subsection{The Safe Interface}

We first consider deriving the safe interface. The new interface
defines a type $\tyalpha~\tau$ corresponding to the base type
$\tau_b$.  The 
instantiations of the
type variable $\tyalpha$ will be used to encode sort
information.  We require an encoding $\langle\sigma\rangle$ of each
sort $\sigma$ in the hierarchy; this encoding should yield a type in
the underlying SML type system, with the property that
$\langle\sigma_1\rangle$ unifies with $\langle\sigma_2\rangle$ if and
only if $\sigma_1$ is a subtype of $\sigma_2$ in the hierarchy.  An
obvious issue is that we want to use unification (a symmetric
relation) to capture subtyping (an asymmetric relation). The simplest
approach is to use two encodings $\conc{\cdot}$ and $\abst{\cdot}$
defined over all the sorts in the hierarchy. A \emph{value}
of sort $\sigma$ will be assigned a type
$\conc{\sigma}~\tau$.  We call $\conc{\sigma}$ the
\emph{concrete} encoding of $\sigma$, and we assume that it
uses only ground types (i.e., no type variables). In order to restrict
the domain of an operation to the set of values 
that are subtypes of a sort $\sigma$,
we use $\abst{\sigma}$, the
\emph{abstract} encoding of $\sigma$.  In order for the
underlying type system to enforce the
subtyping hierarchy, we require the encodings $\conc{\cdot}$ and
$\abst{\cdot}$ to be \emph{respectful} by satisfying the following
property:
\[
\mbox{
for all $\sigma_1$ and $\sigma_2$, 
$\conc{\sigma_1}$ matches $\abst{\sigma_2}$ 
iff $\sigma_1\leq\sigma_2$
}.
\]
For example, the encodings used in the introduction are respectful:
\[
\begin{array}{rclcrcl}
\abst{\itype{atom}} & \teq & \type{$\tyalpha$} & &
\conc{\itype{atom}} & \teq & \type{unit} \\
\abst{\itype{int}} & \teq & \type{int} & &
\conc{\itype{int}} & \teq & \type{int} \\
\abst{\itype{bool}} & \teq & \type{bool} & &
\conc{\itype{bool}} & \teq & \type{bool}.
\end{array}
\]
The utility of the phantom-types technique relies on being able to
find respectful encodings for subtyping hierarchies of interest.

To allow for matching, the abstract encoding will introduce
free type variables. Since, in a Hindley-Milner type system, a type
cannot contain free type variables, the abstract encoding will be part
of the larger type scheme of some polymorphic function operating on
values of appropriate sorts. This leads to some restrictions on
when we should constrain values by concrete or abstract encodings.
For the time being, we will restrict ourselves to using concrete
encodings in all covariant type positions and using abstract
encodings in most contravariant type positions. 
It is fairly easy to see that if we do not impose this restriction, then we
can assign type to functions that break the desired subtyping
invariants. 
For example, suppose we added the following function to our collection
of ``safe'' atom operations:
\begin{center}
\begin{minipage}[t]{0.750\textwidth}
\enablettchars
fun randAtom ():'a safe\_atom =
  (case rand()
    of 0 => W (B (false))
     | 1 => W (B (true))
     | i => W (I (i))).
\end{minipage}
\end{center}
Note that \expr{randAtom} is assigned the type \type{unit -> 'a
safe\_atom}, which appears to be consistent with the fact that
\expr{randAtom} returns some subtype of the sort \itype{atom}.
However, the expression \expr{conj (randAtom (), randAtom ())} is
considered well-typed, but its evaluation may raise a run-time
exception.  Intuitively, the issue with returning a value constrained
by an abstract encoding is that we are trying to impose a restriction
on the behavior of the function based on the types of future uses of
the returned value; such type-directed behavior is not supported in a
language like SML.
We will return to this issue in Section~\ref{s:bounded}.  Another
consequence of having the abstract encoding be part of a larger type
scheme that binds the free variables in prenex position is that the
subtyping is resolved not at the point of function application, but
rather at the point of \emph{type application}, when the type
variables are instantiated. We postpone a discussion of this important
point to Section~\ref{s:bounded}, where we extend our recipe to
account for a form of bounded polymorphism.

\newsavebox{\unsafesigBox}
\begin{lrbox}{\unsafesigBox}
\begin{minipage}[t]{0.350\textwidth}
\scriptsize\enablettchars
signature ATOM = sig
  type atom
  val mkInt: int -> atom
  val mkBool: bool -> atom
  val toString: atom -> string
  val double: atom -> atom
  val conj: atom * atom -> atom
end
\end{minipage}
\end{lrbox}

\newsavebox{\safesigBox}
\begin{lrbox}{\safesigBox}
\begin{minipage}[t]{0.575\textwidth}
\scriptsize\enablettchars
signature SAFE\_ATOM = sig
  type 'a safe\_atom
  val mkInt: int -> \(\conc{\itype{int}}\) safe\_atom
  val mkBool: bool -> \(\conc{\itype{bool}}\) safe\_atom
  val toString: \(\abst{\itype{atom}}\) safe\_atom -> string
  val double: \(\abst{\itype{int}}\) safe\_atom -> \(\conc{\itype{int}}\) safe\_atom
  val conj: \(\abst{\itype{bool}}\) safe\_atom * \(\abst{\itype{bool}}\) safe\_atom 
            -> \(\conc{\itype{bool}}\) safe\_atom
end
\end{minipage}
\end{lrbox}

\begin{figure}
\hrule
\center{
\subfigure[\label{f:unsafesig}Unsafe interface]{%
\usebox{\unsafesigBox}
}\hspace{\fill}%
\subfigure[\label{f:safesig}Safe interface]{%
\usebox{\safesigBox}
}}%
\hrule
\caption{Interfaces for atoms}
\end{figure}

\newsavebox{\impABox}
\begin{lrbox}{\impABox}
\begin{minipage}[t]{0.450\textwidth}
\scriptsize\enablettchars
structure SafeAtom1 :> SAFE\_ATOM = struct
  type 'a safe\_atom = Atom.atom
  val mkInt = Atom.mkInt
  val mkBool = Atom.mkBool
  val toString = Atom.toString
  val double = Atom.double
  val conj = Atom.conj
end
\end{minipage}
\end{lrbox}

\newsavebox{\impBBox}
\begin{lrbox}{\impBBox}
\begin{minipage}[t]{0.500\textwidth}
\scriptsize\enablettchars
structure SafeAtom2 : SAFE\_ATOM = struct
  datatype 'a safe\_atom = W of Atom.atom
  fun int (i) = W (Atom.mkInt (i))
  fun bool (b) = W (Atom.mkBool (b))
  fun toString (W v) = Atom.toString (v)
  fun double (W v) = W (Atom.double (v))
  fun conj (W b1, W b2) = W (Atom.conj (b1,b2))
end
\end{minipage}
\end{lrbox}

\begin{figure}
\hrule
\center{
\subfigure[\label{f:impA}Opaque signature]{%
\usebox{\impABox}
}\hspace{\fill}%
\subfigure[\label{f:impB}Datatype declaration]{
\usebox{\impBBox}
}}%
\hrule
\caption{Two implementations of the safe interface for atoms}
\end{figure}

Consider again the atom example from the introduction.  Assume we have
encodings $\conc{\cdot}$ and $\abst{\cdot}$ for the hierarchy and a structure
\expr{Atom} implementing the ``unsafe'' operations, with the signature
\expr{ATOM}
given in Figure~\ref{f:unsafesig}.  Deriving an interface using the
recipe above, we get the safe signature given in
Figure~\ref{f:safesig}.\footnote{The signature we use is fairly
minimal. There are other functions that would be useful, and that are
still safe with respect to our definition of safety that we use in
Section~\ref{s:formal}. For instance, consider the following function:
\begin{center}
\begin{minipage}[t]{0.850\textwidth}
\enablettchars
fun f (b) = if b then SafeAtom.mkInt (3) else SafeAtom.mkBool (false).
\end{minipage}
\end{center}
This function does not type-check, since \expr{SafeAtom.mkInt (3)} has
type \type{int safe\_atom} while \expr{SafeAtom.mkBool (false)} has
type \type{bool safe\_atom} (assuming the concrete and abstract
encodings defined earlier). What one wants here are 
\emph{coercion functions}, that take values of sort \itype{bool}  or sort
\itype{int} and coerce them to the sort \itype{atom}. This
corresponding to adding a function \expr{coerceToAtom} of type
\type{$\tyalpha$ safe\_atom $\tyto$ unit safe\_atom} to the safe
interface for atoms; the implementation of this function is simply the
identity function. We will not use coercion functions in this
paper, but they can be added without difficulty.}

\subsection{The Safe Implementation}\label{s:safe}

We must now derive an implementation corresponding to the safe signature.  We need a
type $\tyalpha~\tau$ isomorphic to $\tau_b$ such that the type system considers
$\tau_1$ and $\tau_2$ equivalent when $\tau_1~\tau$ and $\tau_2~\tau$ are equivalent.
We can then constrain the types of values and operations using
$\conc{\sigma}~\tau$ and $\abst{\sigma}~\tau$, as indicated above.
There are two ways of enforcing this equivalence in SML:

\begin{enumerate}
\item We can use an abstract type at the module system level, as
shown in Figure~\ref{f:impA}.  The use of an opaque signature is
critical to get the required behavior in terms of type equivalence.
The advantage of this method is that there is no overhead.
\item We can wrap the base type $\tau_b$ using a datatype
declaration
\begin{center}
\begin{minipage}[t]{0.750\textwidth}
\enablettchars
datatype $\tyalpha$ $\tau$ = W of $\tau_b$.
\end{minipage}
\end{center}
The type $\tyalpha~\tau$ behaves as required, because the datatype declaration
defines a generative type operator. However, we must explicitly
convert values of the base type 
to and from $\tyalpha~\tau$ to witness the isomorphism.  This yields the
implementation given in Figure~\ref{f:impB}.
\end{enumerate}
Note that the equivalence requirement precludes the use of type
abbreviations of the form \expr{type $\tyalpha~\tau$ = $\tau_b$}
not restricted by an opaque signature, 
which generally define constant type functions. Moreover, in a
language such as Haskell which does not provide abstract types at the
module level, the second approach is the only one available.
Therefore, for the sake of generality, we use the second approach
throughout this paper, with the understanding that our discussion can
be straightforwardly adapted to the first approach.

We should stress that the safe interface must ensure that the type
$\tyalpha~\tau$ is abstract---either through the use of opaque signature
matching, or by hiding the value constructors of the type. Otherwise,
it may be possible to create values that do not respect the subtyping
invariants enforced by the encodings. 
For example, exposing the wrapper constructor allows a client to write
the following, which type-checks, but violates the implicit subtyping:
\begin{center}
\begin{minipage}[t]{0.750\textwidth}
\enablettchars
val bogus = (W (Atom.mkInt 5)) : bool safe\_atom
val bad = conj (bogus, bogus).
\end{minipage}
\end{center}
The evaluation of \expr{conj (bogus, bogus)} will raise a run-time
exception.
This example demonstrates the subtle difference between the guarantee
made by the SML type-system and the guarantee made by a library
employing the phantom-types technique.  While the former ensures that
``a well-typed program won't go wrong,'' the latter ensures that ``a
well-typed client won't go wrong, provided the library is correctly
implemented.''  The purpose of this section has been to better
characterize what it means for a library to be ``correctly
implemented,'' while Section~\ref{s:formal} will make this
characterization precise.

We now have a way to derive a safe interface and implementation, by
adding type information to a generic, unsafe implementation.  In the
next section, we show how to construct respectful encodings $\conc{\cdot}$
and $\abst{\cdot}$ by taking advantage of the structure of the subtyping
hierarchy.

\section{Encoding Hierarchies}
\label{s:encodings}

The framework presented in the previous section relies on having
concrete and abstract encodings of the sorts in the subtyping
hierarchy with the property that unification of the results of the
encoding respects the subtyping relation.  In
this section, we describe how such encodings can be obtained.
Different encodings are appropriate, depending on the characteristics
of the subtyping hierarchy being encoded.
We assume, for the purpose of this paper, that subtyping hierarchies
are at least join-semilattices. 
These encodings assume that the subtyping relation is completely known
\emph{a priori}. We address the question of extensibility in
Section~\ref{s:extensibility}. 

\subsection{Tree Hierarchies}
\label{s:tree}

For the time being, we restrict ourselves to finite subtyping
hierarchies.  The simplest case to handle is a tree hierarchy.  Since
this is the type of hierarchy that occurs in 
both examples discussed
in the introduction (and, in fact, in all the examples we found in the
literature on encoding subtyping hierarchies in a polymorphic type
system), this encoding should be clear. The idea is to assign a type
constructor to every subtype of a subtyping hierarchy. Assume we have
an encoding $\namesigma{\cdot}$ assigning a distinct (syntactic) name
to each entry in a subtyping hierarchy $(\Sigma,\leq)$.  Hence,
for each $\sigma \in \Sigma$, we define:
\begin{center}
\begin{minipage}[t]{0.750\textwidth}
\enablettchars
datatype $\tyalpha$ $\namesigma{\sigma}$ = Irrelevant\_$\namesigma{\sigma}$.
\end{minipage}
\end{center}
(The name of the data constructor is completely irrelevant, as we will
never construct values of this type.
It is required because SML does not allow the definition of a datatype
with no constructors.)

For example, consider the following subtyping hierarchy (which is
essentially the one used in the Sockets API described by
Reppy~\shortcite{Reppy96}):
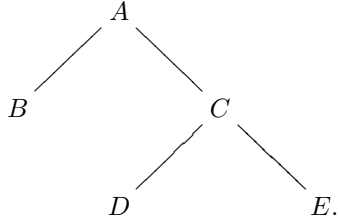
\begin{displaymath}
\begin{xy}
\xymatrix{
& A \ar@{-}[dl] \ar@{-}[dr] & & \\
B & & C \ar@{-}[dl] \ar@{-}[dr] & \\
& D & & E.
}
\end{xy}
\end{displaymath}
We first define type constructors for every element of the hierarchy.
We assume a reasonable name encoding $\namesigma{\cdot}$, such as
$\namesigma{A}=\mbox{\expr{A}}$,
$\namesigma{B}=\mbox{\expr{B}}$, etc.  Hence, we have
\begin{center}
\begin{minipage}[t]{0.750\textwidth}
\enablettchars
datatype $\tyalpha$ A = Irrelevant\_A
datatype $\tyalpha$ B = Irrelevant\_B
\end{minipage}
\end{center}
and likewise for the other elements.  The concrete encoding for an
element of the hierarchy represents the path from the top of the
hierarchy to the element itself. Hence,
\begin{eqnarray*}
\conc{A} & \teq & \mbox{\type{unit A}}\\
\conc{B} & \teq & \mbox{\type{(unit B) A}}\\
\conc{C} & \teq & \mbox{\type{(unit C) A}}\\
\conc{D} & \teq & \mbox{\type{((unit D) C) A}}\\
\conc{E} & \teq & \mbox{\type{((unit E) C) A}}.
\end{eqnarray*}
For the corresponding abstract encoding to be respectful, we require
the abstract encoding of $\sigma$ to unify with the concrete encoding of
all the subtypes of $\sigma$. In other words, we require the abstract
encoding to represent the prefix of the path leading to the element
$\sigma$ in the hierarchy. We use a type variable to unify with any part of
the path after the prefix we want to represent. Hence,
\begin{eqnarray*}
\abst{A} & \teq & \mbox{\type{$\tyalpha_1$ A}}\\
\abst{B} & \teq & \mbox{\type{($\tyalpha_2$ B) A}}\\
\abst{C} & \teq & \mbox{\type{($\tyalpha_3$ C) A}}\\
\abst{D} & \teq & \mbox{\type{(($\tyalpha_4$ D) C) A}}\\
\abst{E} & \teq & \mbox{\type{(($\tyalpha_5$ E) C) A}}.
\end{eqnarray*}
We can then verify, for example, that the concrete encoding of $D$
unifies with the abstract encoding of $C$, as required.

Note that $\abst{\cdot}$ requires every type variable $\tyalpha_i$ to be
a fresh variable, unique in its context.  This ensures that we do not
inadvertently refer to any type variable bound in the context where we
are introducing the abstractly encoded type.  The following example
illustrates the potential problem.  Let $(\Sigma,\leq)$ be the
subtyping hierarchy given above, over some underlying
base
type $\tau_b$.  Suppose we wish to encode an operation 
of type $A \tytimes A \tyto \type{int}$ with the understanding
that a (different) subtype of $A$ may be passed for each of the
arguments. The encoded type of the operation becomes $\abst{A}~\tau
\tytimes \abst{A}~\tau \tyto \type{int}$ (where $\tyalpha~\tau$ is the
wrapped type of $\tau_b$ values) which should translate to
$(\type{$\tyalpha$ A})~\tau \tytimes (\type{$\tybeta$ A})~\tau \tyto
\type{int}$.  If we are not careful in choosing fresh type variables,
we could generate the following type $(\type{$\tyalpha$ A})~\tau
\tytimes (\type{$\tyalpha$ A})~\tau \tyto \type{int}$, corresponding
to a function that requires two arguments of the same type, which is
not the intended meaning.  (The handling of introduced type variables
is somewhat delicate; we address the issue in more detail in
Section~\ref{s:bounded}.)

It should be clear how to generalize the above discussion to concrete
and abstract encodings for arbitrary finite tree hierarchies.  Let
$\top_\Sigma$ correspond to the root of the finite tree hierarchy.  Define an
auxiliary encoding $\auxenc{\cdot}$ which can be used to construct chains
of type constructors.
The encoding $\auxenc{\sigma}$ returns a function expecting the type
to ``attach'' at the end of the chain
\begin{eqnarray*}
\auxenc{\top_\Sigma}~(t) & \teq & t~\namesigma{\top_\Sigma}\\
\auxenc{\sigma}~(t) & \teq & \auxenc{\sigma_\parent}~(t~\namesigma{\sigma})
\quad\mbox{where $\sigma_\parent$ is the parent of $\sigma$}.
\end{eqnarray*}
Thus, in the example above, we have:
\begin{eqnarray*}
\auxenc{A}~(t) & \teq & \type{$t$~A}\\
\auxenc{B}~(t) & \teq & \type{($t$~B)~A}\\
\auxenc{C}~(t) & \teq & \type{($t$~C)~A}\\
\auxenc{D}~(t) & \teq & \type{(($t$~D)~C)~A}\\
\auxenc{E}~(t) & \teq & \type{(($t$~E)~C)~A}.
\end{eqnarray*}
Using this auxiliary encoding, we can define the concrete and abstract
encodings
by supplying the appropriate type:
\begin{eqnarray*}
\conc{\sigma} & \teq & \auxenc{\sigma}~(\type{unit})\\
\abst{\sigma} & \teq & \auxenc{\sigma}~(\tyalpha)
\qquad\mbox{where $\tyalpha$ is fresh.}
\end{eqnarray*}

\subsection{Finite Powerset Lattices}
\label{s:powerset}

Not every subtyping hierarchy of interest is a tree. More general
hierarchies can be used to model multiple interface inheritance in an
object-oriented setting. Let us now examine more general subtyping
hierarchies.  We first consider a particular lattice that will be
useful in our development.  Recall that a lattice is a hierarchy where
every set of elements has both a least upper bound and a greatest
lower bound.  Given a finite set $S$, we let the \emph{powerset
lattice} of $S$ be the lattice of all subsets of $S$, ordered by
inclusion, written $(\wp(S),\subseteq)$. We now exhibit an encoding of
powerset lattices.

Let $n$ be the cardinality of $S$ and assume an ordering $s_1,\dots,s_n$
on the elements of $S$.  We encode subset $X$ of $S$ as an $n$-tuple
type, where the $i^\mathrm{th}$ entry expresses that $s_i \in X$ or $s_i
\not\in X$.  First, we introduce a datatype 
that roughly acts as a Boolean value at the level of types:
\begin{center}
\begin{minipage}[t]{0.750\textwidth}
\enablettchars
datatype $\tyalpha$ z = Irrelevant\_z.
\end{minipage}
\end{center}

The encoding of an arbitrary subset of $S$ is given by:
\begin{eqnarray*}
\conc{X} & \teq & t_1 \tytimes \dots \tytimes t_n
\quad
\mbox{where $t_i\teq\left\{\begin{array}{ll}
                       \type{unit} & \mbox{if $s_i\in X$}\\
                       \type{unit z} & \mbox{otherwise}
                       \end{array}\right.$} \\
\abst{X} & \teq & t_1 \tytimes \dots \tytimes t_n
\quad
\mbox{where $t_i\teq\left\{\begin{array}{ll}
                        \type{$\tyalpha_i$} & \mbox{if $s_i\in X$}\\
                        \type{$\tyalpha_i$~z} &
                        \mbox{otherwise.}
                        \end{array}\right.$}
\end{eqnarray*}
Note that $\abst{\cdot}$ requires every type variable $\tyalpha_i$ to be a fresh
type variable, unique in its context.  This ensures that we do not
inadvertently refer to any type variable bound in the context where we
are introducing the abstractly encoded type.

As an example, consider the powerset lattice of $\{1,2,3,4\}$, which
encodes into a four-tuple. We can verify, for example, that the concrete
encoding for $\{2\}$, namely \type{(unit z $\tytimes$ unit $\tytimes$ unit z $\tytimes$ unit z)},
unifies with the abstract encoding for $\{1,2\}$, namely 
\type{($\tyalpha_1$ $\tytimes$ $\tyalpha_2$ $\tytimes$ $\tyalpha_3$ z $\tytimes$ $\tyalpha_4$ z)}. On the
other hand, the concrete encoding of $\{1,2\}$,
namely \type{(unit $\tytimes$ unit $\tytimes$ unit z $\tytimes$ unit z)},
does not unify with the abstract encoding of $\{2,3\}$,
namely \type{($\tyalpha_1$ z $\tytimes$ $\tyalpha_2$ $\tytimes$ $\tyalpha_3$ $\tytimes$ $\tyalpha_4$ z)}.

\subsection{Embeddings}
\label{s:embedding}

The main reason we introduced powerset lattices is the fact that any
finite hierarchy can be embedded in the powerset lattice of a set $S$.
It is a simple matter, given a hierarchy $\Sigma'$ embedded in a
hierarchy $\Sigma$, to derive an encoding for $\Sigma'$ given an
encoding for $\Sigma$. Let $\embed{\cdot}$ be the injection from
$\Sigma'$ to $\Sigma$ witnessing the embedding and let
$\conc[\Sigma]{\cdot}$ and $\abst[\Sigma]{\cdot}$ be the encodings for
the hierarchy $\Sigma$. Deriving an encoding for $\Sigma'$ simply
involves defining $\conc[\Sigma']{\sigma} \teq
\conc[\Sigma]{\embed{\sigma}}$ and $\abst[\Sigma']{\sigma} \teq
\abst[\Sigma]{\embed{\sigma}}$.  It is straightforward to verify that
if $\conc[\Sigma]{\cdot}$ and $\abst[\Sigma]{\cdot}$ are respectful
encodings, so are $\conc[\Sigma']{\cdot}$ and
$\abst[\Sigma']{\cdot}$. By the result above, this allows us to derive
an encoding for an arbitrary finite hierarchy.

To give an example of embedding, consider the following subtyping
hierarchy to be encoded:
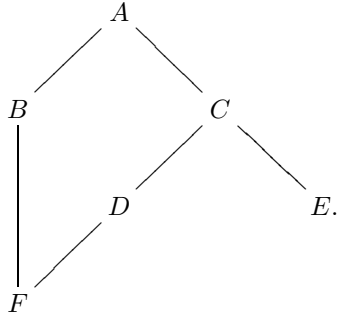
\begin{displaymath}
\begin{xy}
\xymatrix{
& A \ar@{-}[dl] \ar@{-}[dr] & & \\
B \ar@{-}[dd] & & C \ar@{-}[dl] \ar@{-}[dr] & \\
& D \ar@{-}[dl] & & E. \\
F
}
\end{xy}
\end{displaymath}
Notice that this hierarchy can be embedded into the powerset lattice of
$\{1,2,3,4\}$, via the injection function sending $A$ to $\{1,2,3,4\}$,
$B$ to $\{1,2,3\}$, $C$ to $\{2,3,4\}$, $D$ to $\{2,3\}$, $E$ to $\{3,4\}$,
and $F$ to $\{2\}$.

\subsection{Other Encodings}
\label{s:discussion}

We have presented recipes for obtaining respectful encodings,
depending on the characteristics of the 
subtyping 
hierarchy at hand. It should be clear that there are more general
hierarchies than the ones presented here that can still be encoded,
although the encodings quickly become complicated and \emph{ad
hoc}. It would be an interesting project to study in depth the theory
of hierarchies encoding that seems to be lurking here. 
As an example, let us examine an encoding that generalizes
the finite powerset lattice encoding to the (countably) infinite
case, but where only the finite subsets of a countably infinite set
are encoded. Therefore, this encoding is only useful when a program
manipulates only values with sorts corresponding to finite subsets in
the hierarchy. While \emph{ad hoc}, this example is interesting enough
to warrant a discussion. 

Technically, the encoding is in the spirit of the finite powerset
encoding. Let $S$ be a countably infinite set, and assume an ordering
$s_1,s_2,\dots$ of the elements of $S$. As in the finite case, we define a
datatype
\begin{center}
\begin{minipage}[t]{0.750\textwidth}
\enablettchars
datatype $\tyalpha$ z = Irrelevant\_z.
\end{minipage}
\end{center}
The encoding is given for \emph{finite} subsets of $S$ by the
following pair of encodings:
\begin{eqnarray*}
\conc{X} & \teq & (t_1 \tytimes (t_2 \tytimes (t_3 \tytimes \dots \tytimes (t_n \tytimes \tyalpha)\dots))) \\&&
\mbox{where $t_i\teq\left\{\begin{array}{ll}
                       \type{unit} & \mbox{if $s_i\in X$}\\
                       \type{unit z} & \mbox{otherwise}
                       \end{array}\right.$} \\&&
\quad\mbox{and $n$ is the least index such that $X\subseteq\{s_1,\dots,s_n\}$} \\
\abst{X} & \teq & (t_1 \tytimes (t_2 \tytimes (t_3 \tytimes \dots \tytimes (t_n \tytimes \mbox{\type{unit}}),\dots))) \\&&
\mbox{where $t_i\teq\left\{\begin{array}{ll}
                       \type{$\tyalpha_i$} & \mbox{if $s_i\in X$}\\
                       \type{$\tyalpha_i$~z} & \mbox{otherwise}
                       \end{array}\right.$} \\&&
\quad\mbox{and $n$ is the least index such that $X\subseteq\{s_1,\dots,s_n\}$.}
\end{eqnarray*}
(As usual, with the restriction that the type variables
$\tyalpha_1,\dots,\tyalpha_n$ are fresh.) One can verify that
this is indeed a respectful encoding of the finite elements of the
infinite lattice.

Note that this use of a free type variable to be ``polymorphic
in the rest of the encoded value'' is 
strongly reminiscent of the notion of a \emph{row variable}, as
originally used by Wand~\shortcite{Wand87} to type extensible records,
and further developed by R\'{e}my~\shortcite{Remy89}. The technique
was used by Burton~\shortcite{Burton90} to encode extensible records
directly in a polymorphic type system. Recently, the same technique
was used to represents sets of effects in type-and-effect systems~\cite{Pessaux99}.

We have not focussed on the complexity or space-efficiency of the
encodings. We have emphasized simplicity and uniformity of the
encodings, at the expense of succinctness. For instance, deriving an
encoding for a finite hierarchy by embedding it in a powerset lattice
can lead to large encodings even when simpler encodings exist.
Consider the following subtyping hierarchy:
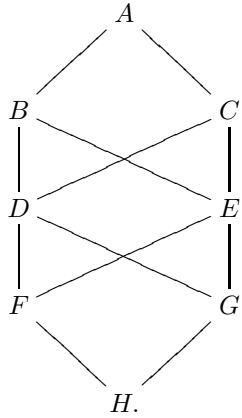
\begin{displaymath}
\begin{xy}
\xymatrix{
& A \ar@{-}[dl] \ar@{-}[dr] & \\
B \ar@{-}[d] \ar@{-}[drr] & & C \ar@{-}[dll] \ar@{-}[d] \\
D \ar@{-}[d] \ar@{-}[drr] & & E \ar@{-}[dll] \ar@{-}[d] \\
F \ar@{-}[dr] & & G \ar@{-}[dl] \\
& H. &
}
\end{xy}
\end{displaymath}
The smallest powerset lattice in which this hierarchy can be embedded
is the powerset lattice of a 6-element set; therefore, the encoding
will require 6-tuples. On the other hand, it is not hard to verify
that the following encoding respects the subtyping induced by this
hierarchy.
As before, we define a datatype
\begin{center}
\begin{minipage}[t]{0.750\textwidth}
\enablettchars
datatype $\tyalpha$ z = Irrelevant\_z.
\end{minipage}
\end{center}
Consider the following encoding:
\begin{eqnarray*}
\conc{A} & \teq & \type{(unit $\tytimes$ unit)} \\
\conc{B} & \teq & \type{(unit z $\tytimes$ unit)} \\
\conc{C} & \teq & \type{(unit $\tytimes$ unit z)} \\
\conc{D} & \teq & \type{((unit z) z $\tytimes$ unit z)} \\
\conc{E} & \teq & \type{(unit z $\tytimes$ (unit z) z)} \\
\conc{F} & \teq & \type{(((unit z) z) z $\tytimes$ (unit z) z)} \\
\conc{G} & \teq & \type{((unit z) z $\tytimes$ ((unit z) z) z)} \\
\conc{H} & \teq & \type{(((unit z) z) z $\tytimes$ ((unit z) z) z)}.
\end{eqnarray*}
The abstract encoding is obtained by replacing every \type{unit} by
a type variable $\tyalpha$, taken fresh, as usual.

It is possible to generate encodings for finite hierarchies that are
in general more efficient than the encodings derived from the powerset
lattice embeddings. One such encoding, 
which we now describe, 
uses a tuple approach just like the
powerset lattice encoding.  This encoding yields tuples whose size
correspond to the width of the subtyping hierarchy being
encoded,
rather than
the typically larger size of the smallest set in whose powerset
lattice the hierarchy can be embedded.  (The efficient encoding for
the previous subtyping hierarchy is an instance of such a width
encoding.)

Let $(\Sigma,\leq)$ be a hierarchy we wish to encode. The
\emph{width} of $\Sigma$ is the maximal size of sets of
incomparable elements. Formally,
\[w(\Sigma) \teq \max\{|X| \mid X\subseteq \Sigma, \forall
x,y\in \Sigma, (x\not\leq y ~\land~ y\not\leq x)\}.\] The
following proposition allows us to derive an encoding based on
the width of the hierarchy. 

\begin{proposition}\label{p:width}
Let $\Sigma$ be a finite hierarchy, and $w$ be the width of
$\Sigma$. There exists  
a function $l:\Sigma\to\mathbb{N}^w$ such that $x\leq y$ if and only if for
$i=1,\dots,w$, $l(x)(i)\geq l(y)(i)$. 
\end{proposition}
\begin{proof}
  Choose $S=\{s_1,\dots,s_w\}$ a subset of $\Sigma$ such that $S$ is a set of
  mutually incomparable elements of size $w$. We iteratively define a
  function $l':\Sigma\to\mathbb{Q}$. We initially set $l'(\top_\Sigma) \teq (0,\dots,0)$,
  and for every $s_i$ in the set $S$,
\[
l'(s_i)  \teq  (a_1,\dots,a_w) \quad 
\mbox{where $a_k\teq\left\{\begin{array}{ll}
                       \frac{1}{2} & \mbox{if $i\not=k$}\\
                       0 & \mbox{otherwise.}
                       \end{array}\right.$}
\]

Iteratively, for all elements $x\in \Sigma$ not assigned a value by $l'$,
define the sets 
\[
x^> \teq \{y \in \Sigma \mid \mbox{$y$ is assigned a value by $l'$ and $y>x$}\}
\] 
and 
\[
x^< \teq \{y\in \Sigma \mid \mbox{$y$ is assigned a value by $l'$ and $y<x$}\}.
\]
It is easy to verify that either $x^>\cap S\not=\emptyset$ or $x^<\cap S\not=\emptyset$, but
not both (otherwise, there exists $y^<\in S$ and $y^>\in S$ such that
$y^>>x>y^<$, and hence $y^>>y^<$, contradicting the mutual
incomparability of elements of $S$). Define $l'(x) \teq (x_1,\dots,x_n)$,
where 
\[
x_i\teq\frac{\min_{y\in x^<}\{l'(y)(i)\} + \max_{y\in x^>}\{l'(y)(i)\}}{2}.
\]

We can now define the function $l:\Sigma\to\mathbb{N}^w$ by simply rescaling
the result of the function $l'$. Let $R$ be the sequence of all the
rational numbers that appear in some tuple position in the result
$l'(x)$ for some $x\in \Sigma$, ordered by the standard order on
$\mathbb{Q}$. For $r\in R$, let $i(r)$ be the index of the rational
number $r$ in $R$. Define the function $l$ by $l(x) \teq
(i(l'(x)(1)),\dots,i(l'(x)(w)))$.  It is straightforward to verify that
the property in the proposition holds for this function. 
\end{proof}

We can use Proposition~\ref{p:width} to encode elements of a finite
hierarchy $\Sigma$. Define a datatype
\begin{center}
\begin{minipage}[t]{0.750\textwidth}
\enablettchars
datatype $\tyalpha$ z = Irrelevant\_z.
\end{minipage}
\end{center}
(As usual, the data constructor name is irrelevant). We encode
an element into a tuple of size $w$, the width of
$\Sigma$. Assume we have a labeling of the elements of
$\Sigma$ by a function $l$ as given by
Proposition~\ref{p:width}.  Essentially, $l$ will indicate the
nesting of the above type constructor in the
encoding. Formally,
\begin{eqnarray*}
\conc{X} & \teq & \underbrace{\type{($\dots$(unit z)$\dots$z)}}_{l(x)(1)} \tytimes \dots \tytimes
               \underbrace{\type{($\dots$(unit z)$\dots$z)}}_{l(x)(w)}\\
\abst{X} & \teq & \underbrace{\type{($\dots$($\tyalpha_1$ z)$\dots$z)}}_{l(x)(1)} \tytimes \dots \tytimes
               \underbrace{\type{($\dots$($\tyalpha_w$ z)$\dots$z)}.}_{l(x)(w)}
\end{eqnarray*}
(As usual, each $\tyalpha_i$ in $\abst{\cdot}$ is fresh.) 

The fact that there are different encodings for the same hierarchy
raises an obvious question: how do we determine the best encoding to
use for a given hierarchy in a given situation? There are interesting
problems here, for instance, lower and upper bounds on optimal
encodings for hierarchies, as well as measurement metrics for
comparing different encodings. We know of no work directly addressing
these issues.

\subsection{Extensibility}
\label{s:extensibility}

One aspect of encodings we have not yet discussed is that of
extensibility. Roughly speaking, extensibility refers to the
possibility of adding new elements to the subtyping hierarchy after a
program has already been written. One would like to avoid having to
rewrite the whole program taking the new subtyping hierarchy into
account. This is especially important in the design of libraries,
where the user may need to extend the kind of data that the library
handles, without changing the provided interface. 
For example, we can easily adapt the subtyping hierarchy of the atom
example to accommodate strings by extending the \expr{SAFE\_ATOM}
signature with
\begin{center}
\begin{minipage}[t]{0.950\textwidth}
\small\enablettchars
val mkString: string -> \(\conc{\itype{str}}\) safe\_atom
val concat: \(\abst{\itype{str}}\) safe\_atom * \(\abst{\itype{str}}\) safe\_atom -> \(\conc{\itype{str}}\) safe\_atom
\end{minipage}
\end{center}
and taking
\[
\begin{array}{rclcrcl}
\abst{\itype{str}} & \teq & \type{string} & &
\conc{\itype{str}} & \teq & \type{string}.
\end{array}
\]
Note that while the implementations of the \expr{Atom} and
\expr{SafeAtom} structures require changes, no existing client of the
\expr{SAFE\_ATOM} signature requires any changes.
In this section, we examine the extensibility of the encodings
we have presented.

Looking at the encodings of Section~\ref{s:encodings}, it should be
clear that the only immediately extensible encodings are the tree
encodings in Section~\ref{s:tree}. In such a case, adding a new
sort $\sigma_\new$ as an immediate subtype of a given sort
$\sigma_\parent$ in the tree simply requires the definition of a new
datatype:
\begin{center}
\begin{minipage}[t]{0.750\textwidth}
\enablettchars
datatype $\tyalpha$ $\namesigma{\sigma_\new}$ = Irrelevant\_$\namesigma{\sigma_\new}$.
\end{minipage}
\end{center}
We assume a naming function $\namesigma{\cdot}$ extended to include $\sigma_\new$.
One can check that the abstract and concrete encodings of the original
elements of the hierarchy are not changed by the extension---since the
encoding relies on the path to the elements.  The concrete and
abstract encodings of the new subtype $\sigma_\new$ is just the path
to $\sigma_\new$, as expected.

The powerset lattice encodings and their embeddings are not so clearly
extensible. 
Indeed, in general, it does not seem possible to arbitrarily extend an
encoding of a subtyping hierarchy 
that contains ``join elements'' (that is, a sort which is a subtype of
at least two otherwise unrelated sorts, related to \emph{multiple
  inheritance} in object-oriented programming). However,  as long as
the extension takes the form of adding new sub-sorts to a single sort
in the hierarchy, it is possible to extend any subtyping hierarchy in
a way that does not invalidate the original  encoding of the
hierarchy. 
As an illustration, consider 
again
the simplest case, where we want to add a new sort $\sigma_\new$ to an
existing hierarchy, where $\sigma_\new$ is 
an immediate
subtype of the sort $\sigma_\parent$. Assume that the existing
hierarchy is encoded using the finite powerset encoding of
Section~\ref{s:powerset}. Observe that in the lattice encodings, the
encoding of a sort $\sigma$ corresponding
to subset $\{s_{i_1},\dots,s_{i_n}\}$ contains a $\mbox{\type{unit}}$
in the tuple positions corresponding to $s_{i_1}$ through $s_{i_n}$, and
$\mbox{\type{unit z}}$ in the other positions. To encode the new sort
$\sigma_\new$, we can simply create a new type
\begin{center}
\begin{minipage}[t]{0.750\textwidth}
\enablettchars
datatype $\tyalpha$ $\namesigma{\sigma_\new}$ = Irrelevant\_$\namesigma{\sigma_\new}$
\end{minipage}
\end{center}
as in the case of tree encodings, and, for the concrete encoding,
replace every $\mbox{\type{unit}}$ in the concrete encoding of
$\sigma_\parent$ by $\mbox{\type{unit $\namesigma{\sigma_\new}$}}$. For
the abstract encoding of $\sigma_\new$, we replace every
$\mbox{\type{unit}}$ in the concrete encoding by a (fresh) type
variable. One can verify that indeed the resulting encoding is
respectful of the subtyping hierarchy with the additional sort
$\sigma_\new$.

We can easily generalize this procedure of adding a new sort to an
existing hierarchy encoded using a powerset lattice encoding to the
case where we want to add a whole hierarchy as subtypes of a single
sort in an existing hierarchy. Here is an outline of the general
approach, of which the above is a special case. 
Let $\Sigma$ be a powerset lattice over a set $S$ of cardinality $n$, and
let $\sigma_\parent$ be an element of $\Sigma$ we want to extend by another
hierarchy $\Sigma_\new$; that is, 
all elements of $\Sigma_\new$ are subtypes of $\sigma_\parent$ and
incomparable to other elements of $\Sigma$.
Assume that $\Sigma$ is encoded via a lattice embedding encoding
$\conc{\cdot}, \abst{\cdot}$, and that $\Sigma_\new$ is encoded via some
encoding $\conc[\new]{\cdot}, \abst[\new]{\cdot}$.  We can extend the encoding
for $\Sigma$ over the elements $\sigma' \in \Sigma_\new$:
\begin{eqnarray*}
\conc{\sigma'} & \teq & t_1 \tytimes \dots \tytimes t_n
\quad
\mbox{where $t_i\teq\left\{\begin{array}{ll}
                       \conc[\new]{\sigma'} & \mbox{if $s_i\in\sigma_\parent$}\\
                       \type{unit z} & \mbox{otherwise}
                       \end{array}\right.$} \\
\abst{\sigma'} & \teq & t_1 \tytimes \dots \tytimes t_n
\quad
\mbox{where $t_i\teq\left\{\begin{array}{ll}
                       \abst[\new]{\sigma'} & \mbox{if $s_i\in\sigma_\parent$}\\
                       \type{$\tyalpha_i$ z} & \mbox{otherwise.}
                       \end{array}\right.$}
\end{eqnarray*}
(As usual, each $\tyalpha_i$ in $\abst{\cdot}$, including the type
variables in $\abst[\new]{\sigma'}$, is fresh.) Again, such an encoding is easily seen
as being respectful of the extended subtyping hierarchy. The above
scheme generalizes in a straightforward way to encodings via lattice
embeddings and to the countable lattice encoding of
Section~\ref{s:discussion}.

As an example, we extend the hierarchy of Section~\ref{s:embedding} as
follows:
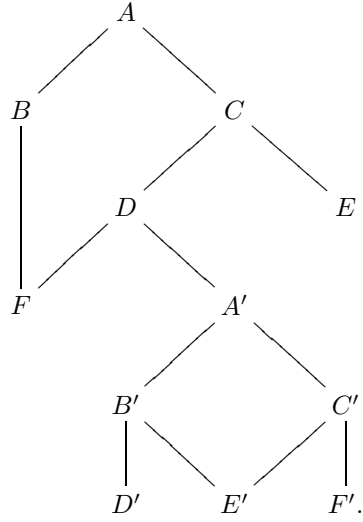
\begin{displaymath}
\begin{xy}
\xymatrix{
& A \ar@{-}[dl] \ar@{-}[dr] & & \\
B \ar@{-}[dd] & & C \ar@{-}[dl] \ar@{-}[dr] & \\
& D \ar@{-}[dl] \ar@{-}[dr] & & E \\
F & & A' \ar@{-}[dl] \ar@{-}[dr] & \\
& B' \ar@{-}[d] \ar@{-}[dr] & & C' \ar@{-}[dl] \ar@{-}[d]\\
& D' & E' & F'.
}
\end{xy}
\end{displaymath}
The complete concrete encoding is given by:
\begin{align*}
\conc[\new]{A'} & \teq  \type{(unit $\tytimes$ unit)} \\
\conc[\new]{B'} & \teq  \type{(unit $\tytimes$ unit z)} \\
\conc[\new]{C'} & \teq  \type{(unit z $\tytimes$ unit)} \\
\conc[\new]{D'} & \teq  \type{(unit $\tytimes$ (unit z) z)} \\
\conc[\new]{E'} & \teq  \type{(unit z $\tytimes$ unit z)} \\
\conc[\new]{F'} & \teq  \type{((unit z) z $\tytimes$ unit)} \\
\end{align*}
and
\begin{align*}
\conc{A} & \teq  \type{(unit $\tytimes$ unit $\tytimes$ unit $\tytimes$ unit)} \\
\conc{B} & \teq  \type{(unit $\tytimes$ unit $\tytimes$ unit $\tytimes$ unit z)} \\
\conc{C} & \teq  \type{(unit z $\tytimes$ unit $\tytimes$ unit $\tytimes$ unit)} \\
\conc{D} & \teq  \type{(unit z $\tytimes$ unit $\tytimes$ unit $\tytimes$ unit z)} \\
\conc{E} & \teq  \type{(unit z $\tytimes$ unit z $\tytimes$ unit $\tytimes$ unit)} \\
\conc{F} & \teq  \type{(unit z $\tytimes$ unit $\tytimes$ unit z $\tytimes$ unit z)} \\[2ex]
\conc{A'} & \teq \type{(unit z $\tytimes$ (unit $\tytimes$ unit) $\tytimes$ (unit $\tytimes$ unit) $\tytimes$ unit z)} \\
\conc{B'} & \teq \type{(unit z $\tytimes$ (unit $\tytimes$ unit z) $\tytimes$ (unit $\tytimes$ unit z) $\tytimes$ unit z)} \\
\conc{C'} & \teq  \type{(unit z $\tytimes$ (unit z $\tytimes$ unit) $\tytimes$ (unit z $\tytimes$ unit) $\tytimes$ unit z)} \\
\conc{D'} & \teq  \type{(unit z $\tytimes$ (unit $\tytimes$ (unit z) z) $\tytimes$ (unit $\tytimes$ (unit z) z) $\tytimes$ unit z)} \\
\conc{E'} & \teq  \type{(unit z $\tytimes$ (unit z $\tytimes$ unit z) $\tytimes$ (unit z $\tytimes$ unit z) $\tytimes$ unit z)} \\
\conc{F'} & \teq  \type{(unit z $\tytimes$ ((unit z) z $\tytimes$ unit) $\tytimes$ ((unit z) z $\tytimes$ unit) $\tytimes$ unit z)}.
\end{align*}
The abstract encoding is obtained by replacing every
\type{unit} by a type variable $\tyalpha$, taken fresh, as usual.

The interesting thing to notice about the above development is that
although extensions are restricted to a single 
sort
(i.e., we can only subtype one given 
sort), 
the extension can itself be an arbitrary lattice. 
As we already pointed out, it does not seem
possible to describe a general extensible encoding that supports
subtyping two different 
sorts 
at the same time (multiple inheritance). In other words, to adopt an
object-oriented perspective, we cannot multiply-inherit from 
multiple sorts
but we can single-inherit into an arbitrary lattice, which can use
multiple inheritance locally.

\section{Encoding a More General Form of Subtyping}
\label{s:bounded}

As mentioned in Section~\ref{s:encodings}, the handling of type
variables is somewhat delicate.  
In this section, we revisit this issue. We show, by approaching the
problem from a different perspective, how we can encode using phantom
types a more general form of subtyping than simply subtyping at
function arguments. We believe that this is the right setting to
understand the ad-hoc restrictions given previously.

If we allow common type variables to be used across abstract
encodings, then we can capture a form of \emph{bounded polymorphism}
as in $\Fsub$~\cite{Cardelli94}.  Bounded polymorphism 
is a typing discipline which extends both parametric
polymorphism and subtyping. From parametric polymorphism, it borrows
type variables and universal quantification; from subtyping, it allows
one to set bounds on quantified type variables.  For example, one can
guarantee that the argument and return types of a function are the
same and a subtype of $\sigma$, as
in $\forall\alpha\leqo\sigma(\alpha\to\alpha)$.\footnote{In
this section, we freely use a $\Fsub$-like notation for
expressions.} Similarly, one can 
guarantee that two arguments have the same type that is a subtype of
$\sigma$, as in
$\forall\alpha\leqo\sigma(\alpha\times\alpha\to\sigma)$. Notice that 
neither function can be written in a language that supports only
subtyping.
In short, bounded polymorphism lets us be more precise when specifying 
subtyping occurring in functions. 

The recipe we gave in Section~\ref{s:subtyping} shows that we can
capture subtyping using parametric polymorphism and restrictions on
type equivalence.  It turns out that we can capture a form of bounded
polymorphism by adapting this procedure. As an example, consider the
type $\forall\beta\leqo\sigma_1(\beta\times\sigma_2\to\beta)$.
Let the ``safe'' interface use types of the form $\alpha~\tau$.  Since
$\beta$ stands for a subtype of $\sigma_1$, we let
$\phi_\beta\teq\abst{\sigma_1}$, the abstract encoding of the bound. We
then translate the type as we did in Section~\ref{s:subtyping}, but
replace occurrences of the type variable $\beta$ by $\phi_\beta$
instead of applying $\abst{\cdot}$ repeatedly. This lets us
\emph{share} the type variables introduced by
$\abst{\sigma_1}$.  Hence, we get the 
type $\phi_\beta~\tau\times\abst{\sigma_2}~\tau\to\phi_\beta~\tau$.
This procedure in fact generalizes that of
Section~\ref{s:subtyping}: we can convert all the subtyping into bounded
polymorphism. More precisely, if a function expects an argument of a
sort that is a
subtype of $\sigma$, we can introduce a fresh type variable for that
argument and bind it by $\sigma$.  For example, the type above can be
rewritten as $\forall\beta\leqo\sigma_1(\forall\gamma\leqo\sigma_2(\beta\times\gamma\to\beta))$,
and encoded as $\phi_\beta~\tau\times\phi_\gamma~\tau\to\phi_\beta~\tau$,
where $\phi_\beta \teq \abst{\sigma_1}$ and $\phi_\gamma \teq \abst{\sigma_2}$.

As one might expect, this technique does not generalize to full
$\Fsub$. For example, it is not clear how to encode types using
bounded polymorphism where the bound 
on a type variable uses a type variable, such as a function \expr{f}
with type
$\forall\alpha\leqo\sigma(\forall\beta\leqo\alpha(\alpha\times\beta\to\alpha))$.
Encoding this type as
$\phi_\alpha~\tau\times\phi_\beta~\tau\to\phi_\alpha~\tau$, where
$\phi_\alpha \teq \abst{\sigma}$ and $\phi_\beta \teq \abst{\alpha}$, fails,
because we have no definition of $\abst{\alpha}$.  Essentially, we
need a different encoding of $\beta$ for each instantiation of
$\alpha$ at each application of \expr{f}, something that cannot be
accommodated by a single encoding of the type at the definition of
\expr{f}.

However, the most important restriction on the kind of bounded
polymorphism that can be handled in a straightforward way is due to
the fact that we are capturing this form of subtyping using SML, which 
uses \emph{prenex} parametric polymorphism. This means, for instance,
that we cannot encode first-class polymorphism, such as a
function \expr{g} with type
$\forall\alpha\leqo\sigma_1(\alpha\to(\forall\beta\leqo\sigma_2(\beta\to\beta)))$.
Applying the technique yields a type
$\phi_\alpha~\tau\to\phi_\beta~\tau\to\phi_\beta~\tau$ where
$\phi_\alpha$ and $\phi_\beta$ contain free type variables.  A
Hindley-Milner style type system requires quantification over these
variables in prenex position, which doesn't match the intuition of the
original type. 
Thus, because we are translating into a language
with prenex polymorphism, it seems we can only capture bounded
polymorphism that is itself in prenex form.

One consequence of being restricted to prenex bounded polymorphism is
that we cannot account for the general subsumption rule found in
$\Fsub$.  Instead, we require all subtyping to occur at type
application.  This is why we can convert all subtyping into bounded
polymorphism, as we did above. By introducing type variables for each
argument, we move the resolution of the subtyping to the point of type
application (when we instantiate the type variables).  The following
example may illustrate this point.  In $\Fsub$ with first-class
polymorphism, we can write a function \expr{app1} with type
$(\forall\alpha\leqo\sigma_1(\alpha\to\sigma_2))\to\sigma_2\times\sigma_2$ that applies a function to two values,
\expr{v1} of type $\sigma_1$ and \expr{v2} of type
$\sigma_2\leq\sigma_1$,
using an SML-like syntax that should be self-explanatory:
\begin{center}
\begin{minipage}[t]{0.750\textwidth}
\enablettchars
local
  val v1 : $\sigma_1$ = $\ldots$
  val v2 : $\sigma_2$ = $\ldots$
in
  fun app1 (f : $\forall\alpha\leq\sigma_1(\alpha\to\sigma_2)$) : $\sigma_2\times\sigma_2$ =
    (f $[\sigma_1]$ v1, f $[\sigma_2]$ v2)
end.
\end{minipage}
\end{center}
This definition of \expr{app1} type-checks when we apply the argument
function to $\sigma_1$ and then to \expr{v1} and we apply the argument
function to $\sigma_2$ (using subsumption at type application) and then to
\expr{v2}.  But, as we argued above, we cannot encode first-class
polymorphism.  An alternative version, \expr{app2}, can be written in
$\Fsub$ with type $(\sigma_1\to\sigma_2)\to\sigma_2\times\sigma_2$:
\begin{center}
\begin{minipage}[t]{0.750\textwidth}
\enablettchars
local
  val v1 : $\sigma_1$ = $\ldots$
  val v2 : $\sigma_2$ = $\ldots$
in
  fun app2 (f : $(\sigma_1\to\sigma_2)$) : $\sigma_2\times\sigma_2$ =
    (f v1, f (v2 : $\sigma_1$))
end.
\end{minipage}
\end{center}
This definition of \expr{app2} type-checks when we apply the argument
function to \expr{v1} and we apply the argument function to \expr{v2}
(using subsumption on \expr{v2}, coercing from $\sigma_2$ to $\sigma_1$).  
Yet, we cannot give any reasonable encoding of \expr{app2} into SML,
because it would require applying the argument function to the
encoding of \expr{v1}, with type $\conc{\sigma_1}~\tau$, and to the encoding
of \expr{v2}, with type $\conc{\sigma_2}~\tau$; that is, it would require
applying an argument function at two different types.  As hinted
above, this is a consequence of the lack of first-class polymorphism
in the SML type system; the argument function cannot be polymorphic.

These two restrictions impose one final restriction on the kind of
subtyping we can encode.  Consider a higher-order function \expr{h}
with type
$\forall\alpha\leqo(\sigma_1\to\sigma_2)(\alpha\to\sigma_2)$.  What
are the possible encodings of the bound $\sigma_1\to\sigma_2$ that
allow subtyping?  Clearly encoding the bound as
$\conc{\sigma_1}~\tau\to\conc{\sigma_2}~\tau$ does not allow any
subtyping.  Encoding the bound as
$\abst{\sigma_1}~\tau\to\abst{\sigma_2}~\tau$ or
$\abst{\sigma_1}~\tau\to\conc{\sigma_2}~\tau$ leads to an unsound
system.  (Consider applying the argument function to a value of type
$\sigma_0\geq\sigma_1$, which would type-check in the encoding,
because $\conc{\sigma_0}$ unifies with $\abst{\sigma_1}$ by the
definition of a respectful encoding.)  However, we can soundly encode
the bound as $\conc{\sigma_1}~\tau\to\abst{\sigma_2}~\tau$.  This
corresponds to a subtyping rule on functional types that asserts
$\tau_1\to\tau_2\leq\tau_1\to\tau'_2$ if and only if
$\tau_2\leq\tau'_2$. 
This is the main reason why we focus on first-order subtyping in this
paper. 

Despite these restrictions, the phantom-types technique is still a
viable method for encoding subtyping in a language like SML.  All of
the examples of phantom types found in the literature satisfy these
restrictions.  In practice, one rarely needs first-class polymorphism
or complicated dependencies between the subtypes of function
arguments, particularly when implementing a safe interface to existing
library functions.

\section{A Formalization}
\label{s:formal}

As the previous section illustrates, there are subtle issues regarding
the kind of subtyping that can be captured using phantom types.  In
this section, we clarify the picture by exhibiting a typed calculus
with a suitable notion of subtyping that can be faithfully translated
into a language such as SML, via a phantom types encoding. The idea is
simple: to see if an interface can be implemented using phantom types,
first express the interface in this calculus in such a way that the
program type-checks.  If it is possible to do so, our results show
that a translation using phantom types exists.  The target of the
translation is a calculus embodying the essence of SML, essentially
the calculus of Damas and Milner~\shortcite{Damas82}, a predicative
polymorphic lambda calculus.

\subsection{The Source Calculus $\scalc$}

\newcommand{\FV}{\mathit{FV}}

Our source calculus, $\scalc$, is a variant of the Damas-Milner
calculus with a restricted notion of bounded polymorphism, and allowing
multiple types for constants. We assume a partially ordered set
$(T,\leq)$ of base types,
which forms the subtyping hierarchy.
\begin{display}{Types of $\scalc$:}
\category{\tau}{types}\\
\entry{t}{base type $(t\in T)$}\\
\entry{\alpha}{type variable}\\
\entry{\tau_1\to\tau_2}{function type}\\
\category{\sigma}{prenex quantified type scheme}\\
\entry{\Forn{\alpha}{\tau}{\tau}}{($\FV(\tau_i)=\emptyseq$, for all $i$)}
\end{display}
Given a type $\tau$ in $\scalc$, we define $\FV(\tau)$ to
be the sequence of type variables appearing in $\tau$, in
depth-first, left-to-right order. (Since there is no
binder in $\tau$, all the type variables appearing in $\tau$ are
necessarily free.) We write sequences using the notation $\langle
\alpha_1,\dots,\alpha_n\rangle$. 
We make a syntactic restriction that precludes the use of type
variables in the bounds of quantified type variables.

An important aspect of our calculus, at least for our purposes, is the
set of
constants that we allow. We distinguish between two types of
constants: base constants and primitive operations. Base constants,
taken from a set $C$, are constants representing values of base types
$t \in T$. We suppose a function $\pi_C:C\to T$ assigning a base type
to every base constant. The primitive operations, taken from a set
$F$, are operations acting on constants and returning
constants.\footnote{For simplicity, we will not deal with higher-order
primitive operations here---they would simply complicate the formalism
without bringing any new insight.  Likewise, allowing primitive
operations to act on and return tuples of values is a simple extension
of the formalism presented here.}  Rather than giving primitive
operations polymorphic types, we assume that the operations can have
multiple types, which encode the allowed subtyping.
We suppose a function $\pi_F$ assigning to every primitive operation
$f\in F$ a set of types $\pi_F(f)$, each type a functional type of the
form $t\to t'$ (for $t,t'\in T$).

Our expression language is a typical polymorphic lambda calculus
expression language.
\begin{renewcommand}{\ratio}{.3}
\begin{display}{Expression Syntax of $\scalc$:}
\category{e}{monomorphic expressions}\\
\entry{c}{base constant $(c\in C)$}\\
\entry{f}{primitive operation $(f\in F)$}\\
\entry{\lam{x}{\tau}{e}}{functional abstraction}\\
\entry{e_1~e_2}{function application}\\
\entry{x}{variable}\\
\entry{p~[\tau_1,\dots,\tau_n]}{type application}\\
\entry{\Let{x}{p}{e}}{local binding}\\
\category{p}{polymorphic expressions}\\
\entry{x}{variable}\\
\entry{\Lamn{\alpha}{\tau}{e}}{type abstraction
($\FV(\tau_i)=\emptyseq$, for all $i$)}
\end{display}
\end{renewcommand}
The operational semantics is given using a standard contextual
reduction semantics, written $e_1\slra e_2$. 
While the details can be found
in Appendix~\ref{a:scalc}, we note here the most important reduction
rule, involving constants:
\[
f~c \slra c' \mbox{~~iff~~} \delta(f,c)=c'
\]
where $\delta: F \times C \rightharpoonup C$ is a partial function
defining the result of applying a primitive operation to a base
constant. 

As previously noted, we do not allow primitive operations to be
polymorphic.  However, we can easily use the fact that they can take
on many types to write polymorphic wrappers.  
For example, we can write a polymorphic wrapper
$\Lam{\alpha}{\tau}{\lam{x}{\alpha}{f~x}}$ to capture the
expected behavior of a function $f$ that may be applied to any subtype
of $\tau$.
We will see shortly that this function is well-typed.

The typing rules for $\scalc$ are the standard Damas-Milner typing
rules, modified to account for subtyping.  The full set of rules is
given in Appendix~\ref{a:scalc}.  Subtyping is given by a judgment
$\Delta\sjudgest{\tau_1}{\tau_2}$ and is derived from the subtyping on the base
types.  The interesting rules are:
\begin{smath}
\Rule{t_1\leq t_2}
     {\Delta\sjudgest{t_1}{t_2}}
\quad\quad
\Rule{\Delta\sjudgest{\tau_2}{\tau'_2}}
     {\Delta\sjudgest{\tau_1\to\tau_2}{\tau_1\to\tau'_2}}.
\end{smath}%
Notice that subtyping at higher types only involves the result type,
following our discussion in 
Sections~\ref{s:subtyping} and \ref{s:bounded}.
The typing rules are given by judgments $\sjudget{\Delta;\Gamma}{e}{\tau}$ for
types and $\sjudget{\Delta;\Gamma}{p}{\sigma}$ for type schemes.  The rule for
primitive operations is interesting:
\begin{smath}
\RuleSide{\text{For all $i$ and for all $\tau_i'$ such that $\sj\tau_i'\subt\tau_i$,}\\
           (\tau'\to\tau)\{\tau'_1/\alpha_1,\dots,\tau'_n/\alpha_n\}\in\pi_F(f)}
         {\Delta,\alpha_1\subt\tau_1,\dots,\alpha_n\subt\tau_n;\Gamma\sj f: \tau'\to\tau}
         {\left(\begin{array}{c}
                f\in F,\\
                \FV(\tau')=\langle\alpha_1,\dots,\alpha_n\rangle
                \end{array}\right).}
\end{smath}%
The syntactic restriction on type variable bounds ensures that each
$\tau_i$ has no type variables, so each $\tau'_i\subt\tau_i$ is well-defined.
The rule captures the notion that any subtyping on a primitive
operation through the use of bounded polymorphism is in fact realized
by the ``many types'' interpretation of the operation.

Subtyping occurs at type application:
\begin{smath}
\Rule{\Delta;\Gamma\sj p:\Forn{\alpha}{\tau}{\tau}
      \quad
      \Delta\sj\tau'_1\subt\tau_1
      \quad
      \dots
      \quad
      \Delta\sj\tau'_n\subt\tau_n}
     {\Delta;\Gamma\sj p~[\tau'_1,\dots,\tau'_n]: \tau\{\tau'_1/\alpha_1,\dots,\tau'_n/\alpha_n\}}.
\end{smath}%
As discussed in the previous section, there is no subsumption in the
system: subtyping must be witnessed by type application. Hence, there
is a difference between the type $t_1\to t_2$ (where $t_1,t_2\in T$) and
$\For{\alpha}{t_1}{\alpha\to t_2}$; namely, the former does not allow any
subtyping.  The restrictions of Section~\ref{s:bounded} are formalized
by prenex quantification and the syntactic restriction on type
variable bounds.

Clearly, type soundness of the above system depends on the definition
of $\delta$ over the constants.  We say that $\pi_F$ is sound with respect to
$\delta$ if for all $f\in F$ and $c\in C$, 
$\sjudget{}{f~c}{\tau}$ implies that $\delta(f,c)$ is defined and
$\pi_C(\delta(f,c))=\tau$.  This definition ensures that any application of
a primitive operation $f$ to a base constant $c$ yields exactly
one value $\delta(f,c)$ at exactly one type $\pi_C(\delta(f,c))=\tau$.  This
leads to the following conditional type soundness result for $\scalc$:
\begin{theorem}\label{t:ssound} 
If $\pi_F$ is sound with respect to $\delta$, $\sjudget{}{e}{\tau}$, and $e\slra
e'$, then $\sjudget{}{e'}{\tau}$ and either $e'$ is a value or there
exists $e''$ such that $e'\slra e''$.
\end{theorem}
\begin{proof} See Appendix~\ref{a:scalc}.\end{proof}

\subsection{The Target Calculus $\tcalc$}

Our target calculus, $\tcalc$, is meant to capture the appropriate
aspects of SML that are relevant for the phantom types encoding of
subtyping. Essentially, it is the Damas-Milner
calculus~\cite{Damas82}
extended with a single type constructor $\mathsf{T}$.
\begin{display}{Types of $\tcalc$:}
\category{\tau}{types}\\
\entry{\alpha}{type variable}\\
\entry{\tau_1\to\tau_2}{function type}\\
\entry{\mathsf{T}~\tau}{type constructor $\mathsf{T}$}\\
\entry{1}{unit type}\\
\entry{\tau_1 \times \tau_2}{product type}\\
\category{\sigma}{prenex quantified type scheme}\\
\entry{\ForT{\alpha_1,\dots,\alpha_n}{\tau}}
\end{display}

\begin{display}{Expression Syntax of $\tcalc$:}
\category{e}{monomorphic expressions}\\
\entry{c}{base constant $(c\in C)$}\\
\entry{f}{primitive operation $(f\in F)$}\\
\entry{\lam{x}{\tau}{e}}{functional abstraction}\\
\entry{e_1~e_2}{function application}\\
\entry{x}{variable}\\
\entry{p~[\tau_1,\dots,\tau_n]}{type application}\\
\entry{\Let{x}{p}{e}}{local binding}\\
\category{p}{polymorphic expressions}\\
\entry{x}{variable}\\
\entry{\LamTn{\alpha}{e}}{type abstraction}
\end{display}

The operational semantics (via a reduction relation $\tlra$) and most
typing rules (via a judgment $\tjudget{\Delta;\Gamma}{e}{\tau}$) are
standard.  The calculus is fully described in Appendix~\ref{a:tcalc}.
As before, we assume that we have 
sets of
constants $C$ and $F$ and a function
$\delta$ providing semantics for primitive applications.  Likewise, we
assume that $\pi_C$ and $\pi_F$ provide types for constants, with
similar restrictions: $\pi_C(c)$ yields a closed type of the form
$\mathsf{T}~\tau$, while $\pi_F(f)$ yields a set of closed types of
the form $(\mathsf{T}~\tau_1)\to(\mathsf{T}~\tau_2)$.  The typing rule
for primitive operations in $\tcalc$ is similar to the corresponding
rule in $\scalc$. 
It ensures that a primitive operation can be given a type (possibly
with free type variables) if all the substitution instances of that
type are allowed by the assignment $\pi_F$. 
Given two types $\tau$ and $\tau'$ in $\tcalc$,
where $\tau'$ is a closed type,
we define their unification $\mathit{unify}(\tau,\tau')$ to be a sequence
of bindings $\langle(\alpha_1,\tau_1),\dots,(\alpha_n,\tau_n)\rangle$ in depth-first, left-to-right order of appearance of
$\alpha_1,\dots,\alpha_n$ in $\tau$
such that $\tau\{\tau_1/\alpha_1,\ldots,\tau_n/\alpha_n\} = \tau'$, 
or $\emptyset$ if $\tau'$ is not a substitution instance of $\tau$.
As for $\scalc$, given a type $\tau$ in $\tcalc$, we define
$\FV(\tau)$ to be the sequence of free type variables appearing in
$\tau$, in depth-first, left-to-right order.
\begin{smath}
\RuleSide{\text{For all $\tau'\in\pi_C(C)$ such that}\\ 
          \mathit{unify}(\tau_1,\tau')=\langle(\alpha_1,\tau'_1),\dots,(\alpha_n,\tau'_n),\dots\rangle,\\
          (\tau_1\to\tau_2)\{\tau'_1/\alpha_1,\dots,\tau'_n/\alpha_n\} \in \pi_F(f)}
         {\Delta,\alpha_1,\dots,\alpha_n;\Gamma\tj f:\tau_1\to\tau_2}
         {\left(\begin{array}{c}
                f\in F,\\
                \FV(\tau_1)=\langle\alpha_1,\dots,\alpha_n\rangle
                \end{array}\right).}
\end{smath}%
Again, this rule captures our notion of ``subtyping through
unification'' by ensuring that the operation is defined at every base
type that unifies with its argument type.  Our notion of soundness of
$\pi_F$ with respect to $\delta$ carries over and we can again establish a
conditional type soundness result: 
\begin{theorem}\label{t:tsound} 
If $\pi_F$ is sound with respect to $\delta$, $\tjudget{}{e}{\tau}$, and $e\tlra
e'$, then $\tjudget{}{e'}{\tau}$ and either $e'$ is a value or there
exists $e''$ such that $e'\tlra e''$.  
\end{theorem}
\begin{proof} See Appendix~\ref{a:tcalc}.\end{proof}

Note that the types $\mathsf{T}~\tau$, $1$, and $\tau_1\times\tau_2$ have no
corresponding introduction and elimination expressions.  We include
these types for the exclusive purpose of constructing the phantom
types used by the encodings.  We could add other types to allow more
encodings, but these suffice for the 
encodings of Section~\ref{s:encodings}.

\subsection{The Translation}
\label{s:translation}

Thus far, we have a calculus $\scalc$ embodying the notion of
subtyping that interests us and a calculus $\tcalc$ capturing the
essence of the SML type system.  We now establish a translation from
the first calculus into the second using phantom types to encode the
subtyping, showing that we can indeed capture that particular notion
of subtyping in SML. Moreover, we show that the translation preserves
the soundness of the types assigned to constants, thereby guaranteeing
that if the original system was sound, the system obtained by
translation is sound as well.

We first describe how to translate types in $\scalc$.  Since subtyping
is only witnessed at type abstraction, the type translation realizes
the subtyping using the phantom types encoding of abstract and
concrete subtypes.  The translation is parameterized by an environment
$\rho$ associating every (free) type variable with a type in $\tcalc$
representing the abstract encoding of the bound.
\begin{display}{Types Translation:}
\clause{\cT\intension{\alpha}\rho \teq \rho(\alpha)}\\
\clause{\cT\intension{t}\rho \teq \mathsf{T}~\conc{t}}\\
\clause{\cT\intension{\tau_1\to\tau_2}\rho \teq
  \cT\intension{\tau_1}\rho\to\cT\intension{\tau_2}\rho}\\
\clause{\cT\intension{\Forn{\alpha}{\tau}{\tau}}\rho \teq
\begin{prog}
\ForT{\alpha_{11},\dots,\alpha_{1k_1},\dots,\alpha_{n1},\dots,\alpha_{nk_n}}{\cT\intension{\tau}\rho[\alpha_i\mapsto\tau_i^A]}\\
\quad \mbox{where $\tau_i^A=\cA\intension{\tau_i}$}\\
\quad \mbox{and $\FV(\tau_i^A)=\langle \alpha_{i1},\dots,\alpha_{ik_i}\rangle$}
\end{prog}}
\end{display}
If $\rho$ is empty, we simply write $\cT\intension{\tau}$
for $\cT\intension{\tau}\rho$. 
To compute the abstract and concrete encodings of a type, we define
the functions $\cA$ and $\cC$. 
\begin{display}{Abstract and Concrete Encodings:}
\clause{\cA\intension{t} \teq \mathsf{T}~\abst{t}}\\
\clause{\cA\intension{\tau_1\to\tau_2} \teq
  \cC\intension{\tau_1} \to \cA\intension{\tau_2}}\\
\clause{\cC\intension{t} \teq \mathsf{T}~\conc{t}}\\
\clause{\cC\intension{\tau_1\to\tau_2} \teq
  \cC\intension{\tau_1} \to \cC\intension{\tau_2}}
\end{display}
The syntactic restriction on type variable bounds ensures
that $\cA$ and $\cC$ are always well defined, as they are never 
applied to type variables.  Furthermore, the above
translation depends on the fact that the type encodings $\conc{t}$ and
$\abst{t}$ are expressible in the $\tcalc$ type system using
$\mathsf{T}$, $1$, and $\times$.

We extend the type transformation $\cT$ to type contexts $\Gamma$ in the
obvious way:
\begin{display}{Type Contexts Translation:}
\clause{\cT\intension{x_1:\tau_1,\dots,x_n:\tau_n}\rho \teq
x_1:\cT\intension{\tau_1}\rho,\dots,x_n:\cT\intension{\tau_n}\rho}
\end{display}

Finally, if we take the base constants and the primitive operations
in $\scalc$ and assume that $\pi_F$ is sound with respect to $\delta$, then
the translation can be used to assign types to the constants and
operations such that they are sound in the target calculus.  We first
extend the definition of $\cT$ to $\pi_C$ and $\pi_F$ in the
obvious way:
\clearpage
\begin{display}{Interpretations Translation:}
\clause{\cT\intension{\pi_C} \teq \pi'_C \quad \mbox{where
$\pi'_C(c) = \cT\intension{\pi_C(c)}$}}\\
\clause{\cT\intension{\pi_F} \teq \pi'_F
\quad \mbox{where $\pi'_F(f) = \{\cT\intension{\tau} \mid \tau\in\pi_F(f)\}$}}
\end{display}
We can further show that the translated types do not allow us to
``misuse'' the constants in $\tcalc$: 
\begin{theorem}\label{t:safe} 
If $\pi_F$ is sound with respect to $\delta$ in $\scalc$, then
$\cT\intension{\pi_F}$ is sound with respect to $\delta$ in $\tcalc$.  
\end{theorem}
\begin{proof} See Appendix~\ref{a:translation}.\end{proof}

We therefore take $\cT\intension{\pi_C}$ and $\cT\intension{\pi_F}$ to be
the interpretations in 
the
target calculus $\tcalc$.

We can now define the translation of expressions via a translation of
typing derivations, $\cE$, taking care to respect the types given by
the above type translation.  We note that the translation below
works only if the concrete encodings being used do not contain free type
variables.  Again, the translation is parameterized by an environment
$\rho$, as in the type translation.
\begin{display}{Expressions Translation:}
\clause{\cE\intension{\sjudget{\Delta;\Gamma}{x}{\tau}}\rho \teq x}\\
\clause{\cE\intension{\sjudget{\Delta;\Gamma}{c}{\tau}}\rho \teq c}\\
\clause{\cE\intension{\sjudget{\Delta;\Gamma}{f}{\tau}}\rho \teq f}\\
\clause{\cE\intension{\sjudget{\Delta;\Gamma}{\lam{x}{\tau'}{e}}{\tau}}\rho
  \teq \lam{x}{\cT\intension{\tau'}\rho}{\cE\intension{e}\rho}}\\
\clause{\cE\intension{\sjudget{\Delta;\Gamma}{e_1~e_2}{\tau}}\rho
  \teq (\cE\intension{e_1}\rho)~\cE\intension{e_2}\rho}\\
\clause{\cE\intension{\sjudget{\Delta;\Gamma}{\Let{x}{p}{e}}{\tau}}\rho
  \teq \Let{x}{\cE\intension{p}\rho}{\cE\intension{e}\rho}}\\
\clause{
  \begin{prog}
  \cE\intension{\sjudget{\Delta;\Gamma}{p~[\tau_1,\dots,\tau_n]}{\tau}}\rho \teq\\
  \quad(\cE\intension{p}\rho)~[\tau_{11},\dots,\tau_{1k_1},\dots,\tau_{n1},\dots,\tau_{nk_n}]\\
  \qquad\mbox{where $\cB\intension{p}\Gamma =
    \langle(\alpha_1,\tau_1^B),\dots,(\alpha_n,\tau_n^B)\rangle$ and $\tau_i^A =
    \cA\intension{\tau_i^B}$}\\
  \qquad\mbox{and $\FV(\tau_i^B) = \langle\alpha_{i1},\dots,\alpha_{ik_i}\rangle$ 
      and $\tau_i^T = \cT\intension{\tau_i}\rho$}\\
  \qquad\mbox{and $\mathit{unify}(\tau_i^A,\tau_i^T) = 
           \langle(\alpha_{i1},\tau_{i1}),\dots,(\alpha_{ik_i},\tau_{ik_i}),\dots\rangle$}
  \end{prog}}\\
\clause{\cE\intension{\sjudget{\Delta;\Gamma}{x}{\sigma}}\rho \teq x}\\
\clause{
  \begin{prog}
  \cE\intension{\sjudget{\Delta;\Gamma}{\Lamn{\alpha}{\tau}{e}}{\sigma}}\rho \teq\\ 
  \quad\LamT{\alpha_{11},\dots,\alpha_{1k_1},\dots,\alpha_{n1},\dots,\alpha_{nk_n}}{\cE\intension{e}\rho[\alpha_i\mapsto\tau_i^A]}\\ 
  \qquad\mbox{where $\tau_i^A=\cA\intension{\tau_i}$ 
      and $\FV(\tau_i^A)=\langle\alpha_{i1},\dots,\alpha_{ik_i}\rangle$}
  \end{prog}}
\end{display}
Again, if $\rho$ is empty, we simply write $\cE\intension{e}$
for $\cE\intension{e}\rho$.  
The function $\cB$ returns the bounds of a type abstraction, using the
environment $\Gamma$ to resolve variables.
\begin{display}{Bounds of a Type Abstraction:}
\clause{\cB\intension{x}\Gamma \teq 
  \langle(\alpha_1,\tau_1),\dots,(\alpha_n,\tau_n)\rangle
  \quad\mbox{where $\Gamma(x)=\Forn{\alpha}{\tau}{\tau}$}}\\
\clause{\cB\intension{\Lamn{\alpha}{\tau}{e}}\Gamma \teq
\langle(\alpha_1,\tau_1),\dots,(\alpha_n,\tau_n)\rangle}
\end{display}
We use $\cB$ and $\mathit{unify}$ to perform unification ``by hand.''
In most programming languages, type inference performs this
automatically.

We can verify that this translation is 
type-preserving:
\begin{theorem}\label{t:correct} 
If $\sjudget{}{e}{\tau}$, then
$\tjudget{}{\cE\intension{\sjudget{}{e}{\tau}}}{\cT\intension{\tau}}$.
\end{theorem}
\begin{proof} See Appendix~\ref{a:translation}.\end{proof}

Theorem~\ref{t:correct} shows that the translation captures the
right notion of subtyping, in the sense that interests us,
particularly when designing an interface.  Given a set of constants
making up the interface, suppose we can assign types to those
constants in $\scalc$ in a way that gives the desired subtyping; that
is, we can write type-correct expressions of the form
$\Lam{\alpha}{t}{\lam{x}{\alpha}{f~x}}$ with type $\For{\alpha}{t}{\alpha\to\tau}$.  In other
words, the typing $\pi_F$ is sound with respect to the semantics of $\delta$.
By Theorem~\ref{t:ssound}, this means that $\scalc$ with these
constants is sound and we can safely use these constants in $\scalc$.
In particular, we can write the program:
\[
\begin{array}{l}
\Let{g_1}{\Lam{\alpha}{t_{i_1}}{\lam{x}{\alpha}{f_1~x}}}{}\\
\quad\vdots\\
\Let{g_n}{\Lam{\alpha}{t_{i_n}}{\lam{x}{\alpha}{f_n~x}}}{}\\
\quad e.
\end{array}
\]
By Theorem~\ref{t:correct}, the translation of the above program
executes without run-time errors.  Furthermore, by
Theorem~\ref{t:safe}, the phantom-types encoding of the types of these
constants are sound with respect to $\delta$ in $\tcalc$.  Hence, by
Theorem~\ref{t:tsound}, $\tcalc$ with these constants is sound and we
can safely use these constants in $\tcalc$.  Therefore, we can replace
the body of the translated program with an arbitrary $\tcalc$
expression that type-checks in that context and the resulting program
will still execute without run-time errors.  Essentially, the
translation of the let bindings corresponds to a ``safe'' interface to
the primitives; programs that use this interface in a type-safe manner
are guaranteed to execute without run-time errors.

\subsection{Example and Remarks}
\label{s:remarks}

In this section, we work through a mostly complete example before
turning our attention to some general remarks.

Recall the subtyping hierarchy introduced in
Section~\ref{s:subtyping}
and here extended to include natural numbers and strings.
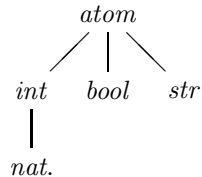
\begin{center}
\setlength{\unitlength}{0.2in}
\begin{picture}(8,6)(0,2)
\put(4,7){\makebox(0,0){\itype{atom}}}
\put(2,5){\makebox(0,0){\itype{int}}}
\put(2,3){\makebox(0,0){\itype{nat}.}}
\put(4,5){\makebox(0,0){\itype{bool}}}
\put(6,5){\makebox(0,0){\itype{str}}}
\put(4.5,6.5){\line(1,-1){1}}
\put(4,6.5){\line(0,-1){1}}
\put(3.5,6.5){\line(-1,-1){1}}
\put(2,4.5){\line(0,-1){1}}
\end{picture}
\end{center}
We can encode this hierarchy with phantom types as follows:
\[
\begin{array}{rclcccrcl}
\abst{\itype{atom}} & = & \alpha \times (\beta \times \gamma) & & & & 
\conc{\itype{atom}} & = & 1 \times (1\times 1) \\
\abst{\itype{int}} & = & \mathsf{T}~\alpha \times (\beta \times \gamma) & & & & 
\conc{\itype{int}} & = & \mathsf{T}~1 \times (1 \times 1) \\
\abst{\itype{nat}} & = & \mathsf{T}~(\mathsf{T}~\alpha) \times (\beta \times \gamma) & & & & 
\conc{\itype{nat}} & = & \mathsf{T}~(\mathsf{T}~1) \times (1 \times 1) \\
\abst{\itype{bool}} & = & \alpha \times (\mathsf{T}~\beta \times \gamma) & & & & 
\conc{\itype{bool}} & = & 1 \times (\mathsf{T}~1 \times 1) \\
\abst{\itype{str}} & = & \alpha \times (\beta \times \mathsf{T}~\gamma) & & & & 
\conc{\itype{str}} & = & 1 \times (1 \times \mathsf{T}~1).
\end{array}
\]
We consider two primitive operations \expr{double} and \expr{toString}
with
\begin{eqnarray*}
\pi_F(\expr{double}) & = &
\{\itype{int}\to\itype{int},~
 \itype{nat}\to\itype{nat}\} \\
\pi_F(\expr{toString}) & = &
\{\itype{atom}\to\itype{str},~
 \itype{int}\to\itype{str},~
 \itype{nat}\to\itype{str},~
 \itype{bool}\to\itype{str},~
 \itype{str}\to\itype{str}\}.
\end{eqnarray*}
We can derive the following typing judgments in $\scalc$, which
capture the intended subtyping:
\[
\begin{array}{l}
\sjudget{}
        {\Lam{\alpha}{\itype{int}}{\lam{x}{\alpha}{\expr{double}~x}}}
        {\For{\alpha}{\itype{int}}{\alpha\to\alpha}} \\
\sjudget{}
        {\Lam{\alpha}{\itype{atom}}{\lam{x}{\alpha}{\expr{toString}~x}}}
        {\For{\alpha}{\itype{atom}}{\alpha\to\expr{str}}.}
\end{array}
\]
Applying our translation to these functions yields:
\begin{eqnarray*}
\cE\intension{\Lam{\alpha}{\itype{int}}{\lam{x}{\alpha}{\expr{double}~x}}} & = &
\LamT{\alpha,\beta,\gamma}{\lam{x}{\mathsf{T}~(\mathsf{T}~\alpha \times (\beta \times\gamma))}{\expr{double}~x}} \\
\cE\intension{\Lam{\alpha}{\itype{atom}}{\lam{x}{\alpha}{\expr{toString}~x}}} & = &
\LamT{\alpha,\beta,\gamma}{\lam{x}{\mathsf{T}~(\alpha \times (\beta \times \gamma))}{\expr{toString}~x}}.
\end{eqnarray*}
As expected from Theorem~\ref{t:correct}, we can derive typing
judgments that assign the translated types to these functions:
\begin{align*}
& \tj \LamT{\alpha,\beta,\gamma}{\lam{x}{\mathsf{T}~(\mathsf{T}~\alpha \times (\beta \times \gamma))}{\expr{double}~x}}:\\
& \qquad\qquad \ForT{\alpha,\beta,\gamma}{\mathsf{T}~(\mathsf{T}~\alpha \times (\beta \times \gamma))\to\mathsf{T}~(\mathsf{T}~\alpha \times (\beta \times \gamma))}\\
& \tj \LamT{\alpha,\beta,\gamma}{\lam{x}{\mathsf{T}~(\alpha \times
(\beta \times \gamma))}{\expr{toString}~x}}:\\
& \qquad\qquad \ForT{\alpha,\beta,\gamma}{\mathsf{T}~(\alpha \times (\beta \times \gamma))\to\mathsf{T}~(1 \times (1 \times \mathsf{T}~1))}.
\end{align*}

Interestingly, we can also derive the following typing judgments:
\begin{align*}
& \tj \LamT{\alpha}{\lam{x}{\mathsf{T}~(\mathsf{T}~\alpha \times (\alpha \times \alpha))}{\expr{double}~x}}\\
& \qquad\qquad \ForT{\alpha}{\mathsf{T}~(\mathsf{T}~\alpha \times(\alpha \times \alpha))\to\mathsf{T}~(\mathsf{T}~\alpha\times (\alpha \times \alpha))}\\
& \tj \LamT{\alpha,\beta}{\lam{x}{\mathsf{T}~(\alpha \times (\beta \times \beta))}{\expr{toString}~x}}\\
& \qquad\qquad \ForT{\alpha,\beta}{\mathsf{T}~(\alpha \times (\beta \times \beta))\to\mathsf{T}~(1 \times (1\times\mathsf{T}~1))}.
\end{align*}
The first function type-checks because, of all base types, only
$\mathsf{T}~\conc{\itype{int}}$ unifies with $\mathsf{T}~(\mathsf{T}~\alpha
\times (\alpha \times \alpha))$, by the substitution $(\alpha,1)$, and
$\{\mathsf{T}~\conc{\itype{int}}\to\mathsf{T}~\conc{\itype{int}}\}\subseteq\cT\intension{\pi_F}(\expr{double})$.
Likewise, the second function type-checks because, of all base types,
only $\mathsf{T}~\conc{\itype{atom}}$,
$\mathsf{T}~\conc{\itype{int}}$, $\mathsf{T}~\conc{\itype{nat}}$ unify
with $\mathsf{T}~(\mathsf{T}~\alpha \times (\beta \times \beta))$ and
$\{\mathsf{T}~\conc{\itype{atom}}\to\mathsf{T}~\conc{\itype{str}},~\mathsf{T}~\conc{\itype{int}}\to\mathsf{T}~\conc{\itype{str}},~\mathsf{T}~\conc{\itype{nat}}\to\mathsf{T}~\conc{\itype{str}}\}\subseteq\cT\intension{\pi_F}(\expr{toString})$.
We can interpret the first as a function that can only be applied to
integers (but not naturals) and the second as a function that can only
be applied to atoms, integers, and naturals (but not booleans or
strings).  Observe that while these functions do not capture all of
the subtyping available in their wrapped primitive operations, neither
do they violate the subtyping available.  This corresponds to the fact
that the second set of types are instances of the first set of types
under appropriate substitutions for $\beta$ and $\gamma$.

The existence of these typing judgments sheds some light on the
practical aspects of using the phantom-types technique in real
programming languages.  Recall that the typing judgment for primitive
operations is somewhat non-standard.  Specifically, in contrast to
most typing judgments for primitives (like the typing judgment for
base constants), this judgment is not syntax directed; that is, the
type is not uniquely determined by the primitive operation.  This
complicates a type-inference system for $\tcalc$.  At the same time,
we cannot expect to integrate this typing judgment into an existing
language with a Hindley-Milner style type system.  Rather, we expect
to integrate a primitive operation into a programming language through
a foreign-function interface, at which point we give the introduced
function a very base type that does not reflect the subtyping
inherent in its semantics.\footnote{In general, foreign-function
  interfaces have strict requirements on the types of foreign
  functions that can be called.  Due to internal implementation
  details, language implementations rarely allow foreign functions to
  be given polymorphic types or types with user defined datatypes,
  both of which are used by the phantom types encodings.}  After
introducing the primitive operation in this fashion, we wrap it with a
function to which we can assign the intended type using the phantom
types encoding, because the type system will not, in general, infer
the appropriate type.  It is for this reason that we have stressed the
application of phantom-types technique to developing and implementing
interfaces.

\section{Conclusion}

The phantom-types technique uses the definition of type equivalence in
a programming language such as SML or Haskell to encode information in
a free type variable of a type.  Unification can then be used to
enforce a particular structure on the information carried by two such
types.  In this paper, we have focused on encoding subtyping
information.  We were able to provide encodings for hierarchies with
various characteristics, 
and more generally, hinted at a theory for how such encodings can be
derived.
Because the technique relies on encoding the subtyping hierarchy, the
problem of extensibility arises: how resilient are the encodings to
additions to the subtyping hierarchy?  This is especially important
when designing library interfaces. We showed in this paper that our
encodings can handle extensions to the subtyping hierarchy as long as
the extensions are always made with respect to a single parent in the
hierarchy.  We also showed how to extend the techniques we developed
to encode a form of prenex bounded polymorphism, with subsumption
occurring only at type application.  The correctness of this encoding
is established by showing how a calculus with that form of subtyping
can be translated faithfully (using the encoding) into a calculus
embodying the type system of SML.

It goes without saying that this approach to encoding subtyping is not
without its problems from a practical point of view. As the encodings
in this paper show, the types involved can become quite large. Type
abbreviations can help simplify the presentation of concrete types,
but abstract encodings require type variables and those 
variables need to appear in the interface. Having such complex types
lead to interfaces themselves becoming complex, and, more seriously,
the type errors reported to the user are fairly unreadable. Although
the process of encoding the subtyping hierarchies can be
automated---for instance, by deriving the encodings from a
declarative description of the hierarchy---we see no good
solution for the complexity problem. The 
compromise between providing safety and complicating the interface
must be decided on a per-case basis.

We also note that the source language of Section~\ref{s:formal}
provides only a lower bound on the power of phantom types.  For
example, one can use features of the \emph{specific} encoding used to
further constrain or refine the type of operations. 
This is used, for instance, 
by Reppy~\shortcite{Reppy96} to type socket operations.  There is yet
no general methodology for exploiting properties of encodings beyond
them respecting the subtyping hierarchy.

\subsection*{Acknowledgments}

A preliminary version of this paper appeared in the
\textit{Proceedings of the 2nd IFIP International Conference on
Theoretical Computer Science}, pp. 448--460, 2002. This work 
was mostly done when the second author was at Cornell University.
We have benefited
from discussions with Greg Morrisett and Dave MacQueen. John Reppy
pointed out the work of Burton. Stephanie Weirich, Vicky Weissman, and
Steve Zdancewic provided helpful comments on an early draft of this
paper.  
Thanks also to the anonymous
referees for many suggestions to improve the presentation.  The second
author was partially supported by ONR grant N00014-00-1-03-41.

\appendix

\renewcommand{\ratio}{.3}

\section{The Calculus $\scalc$}
\label{a:scalc}

\begin{display}{Values:}
\category{v}{values}\\
\entry{c}{base constant $(c\in C)$}\\
\entry{f}{primitive operation $(f\in F)$}\\
\entry{\lam{x}{\tau}{e}}{functional abstraction}
\end{display}

\begin{display}{Evaluation Contexts:}
\category{E}{evaluation contexts}\\
\entry{[~]}{empty context}\\
\entry{E~e}{application context}\\
\entry{v~E}{argument context}\\
\entry{E~[\tau_1,\dots,\tau_n]}{type application context}
\end{display}

\begin{display}{Operational Semantics:}
\clause{(\lam{x}{\tau}{e})~v \slra e\{v/x\}}\\
\clause{(\Lamn{\alpha}{\tau}{e})~
       [\tau'_1,\dots,\tau'_n] \slra
       e\{\tau'_1/\alpha_1,\dots,\tau'_n/\alpha_n\}} \\
\clause{\Let{x}{p}{e} \slra  e\{p/x\}}\\
\clause{f~c \slra  c' \mbox{~~iff~~} \delta(f,c)=c'}\\
\clause{E[e_1]\slra E[e_2] \mbox{~~iff~~} e_1\slra e_2}
\end{display}
The function $\delta: F \times C \rightharpoonup C$ is a partial function defining the
result of applying a primitive operation to a base constant.

\begin{display}{Typing Contexts:}
\category{\Gamma}{type environments}\\
\entry{\cdot}{empty}\\
\entry{\Gamma,x:\tau}{type}\\
\entry{\Gamma,x:\sigma}{type scheme}\\
\category{\Delta}{subtype environments}\\
\entry{\cdot}{empty}\\
\entry{\Delta,\alpha\subt\tau}{subtype}
\end{display}

\begin{display}{Judgments:}
\clause{\sj \Delta~\mbox{ctxt}}{good context $\Delta$}\\
\clause{\Delta\sj \tau~\mbox{type}}{good type $\tau$}\\
\clause{\Delta\sj \sigma~\mbox{scheme}}{good type scheme $\sigma$}\\
\clause{\Delta\sj \Gamma~\mbox{ctxt}}{good context $\Gamma$}\\
\clause{\Delta\sj \tau_1\subt\tau_2}{type $\tau_1$ subtype of $\tau_2$}\\
\clause{\Delta;\Gamma\sj e:\tau}{good expression $e$ with type $\tau$}\\
\clause{\Delta;\Gamma\sj p:\sigma}{good expression $p$ with type scheme $\sigma$}
\end{display}

\begin{display}{Judgment {\mdseries $\sj \Delta~\mbox{ctxt}$}:}
\clause{
\Rule{}{\sj \cdot~\mbox{ctxt}} \quad
\Rule{\sj \Delta~\mbox{ctxt} \quad
      \Delta \sj \tau~\mbox{type}}
     {\sj \Delta,\alpha\subt\tau~\mbox{ctxt}}}
\end{display}

\begin{display}{Judgment {\mdseries $\Delta\sj \tau~\mbox{type}$}:}
\clause{
\Rule{}{\Delta \sj t~\mbox{type}} \quad
\Rule{\sj \Delta~\mbox{ctxt} \quad
      \alpha \in \mbox{dom}(\Delta)}
     {\Delta \sj \alpha~\mbox{type}} \quad
\Rule{\Delta \sj \tau_1~\mbox{type} \quad
      \Delta \sj \tau_2~\mbox{type}}
     {\Delta \sj \tau_1\to\tau_2~\mbox{type}}}
\end{display}

\begin{display}{Judgment {\mdseries $\Delta\sj \sigma~\mbox{scheme}$}:}
\clause{
\Rule{\Delta,\alpha_1\subt\tau_1,\dots,\alpha_n\subt\tau_n \sj \tau~\mbox{type}}
     {\Delta \sj \Forn{\alpha}{\tau}{\tau}~\mbox{scheme}}}
\end{display}

\begin{display}{Judgment {\mdseries $\Delta\sj \Gamma~\mbox{ctxt}$}:}
\clause{
\Rule{}{\Delta\sj\cdot~\mbox{ctxt}} \quad 
\Rule{\Delta\sj\Gamma~\mbox{ctxt} \quad
      \Delta\sj\tau~\mbox{type}}
     {\Delta\sj\Gamma,x:\tau~\mbox{ctxt}} \quad 
\Rule{\Delta\sj\Gamma~\mbox{ctxt} \quad
      \Delta\sj\sigma~\mbox{scheme}}
     {\Delta\sj\Gamma,x:\sigma~\mbox{ctxt}}}
\end{display}

\begin{display}{Judgment {\mdseries $\Delta\sj \tau_1\subt\tau_2$}:}
\clause{
\Rule{}{\Delta\sj \tau\subt\tau} \quad 
\Rule{}{\Delta,\alpha\subt\tau\sj\alpha\subt\tau} \quad 
\Rule{t_1\leq t_2}{\Delta\sj t_1\subt t_2}}\\[\GAP]
\clause{
\Rule{\Delta\sj\tau_2\subt\tau_3}
     {\Delta\sj\tau_1\to\tau_2\subt\tau_1\to\tau_3} \quad
\Rule{\Delta\sj\tau_1\subt\tau_2\quad\Delta\sj\tau_2\subt\tau_3}
     {\Delta\sj\tau_1\subt\tau_3}
}
\end{display}

\begin{display}{Judgment {\mdseries $\Delta;\Gamma\sj e:\tau$}:}
\clause{
\RuleSide{}{\Delta;\Gamma\sj c:\pi_C(c)}{(c\in C)}}\\[\GAP]
\clause{
\RuleSide{\text{For all $i$, and for all $\tau_i'$ such that $\sj\tau_i'\subt\tau_i$,}\\
           (\tau'\to\tau)\{\tau'_1/\alpha_1,\dots,\tau'_n/\alpha_n\}\in\pi_F(f)}
         {\Delta,\alpha_1\subt\tau_1,\dots,\alpha_n\subt\tau_n;\Gamma\sj f: \tau'\to\tau}
         {\left(\begin{array}{c}
                f\in F,\\
                \FV(\tau')=\langle\alpha_1,\dots,\alpha_n\rangle
                \end{array}\right)}
}\\[\GAP]
\clause{
\Rule{\Delta\sj\Gamma~\mbox{ctxt}}{\Delta;\Gamma,x:\tau\sj x:\tau} \quad 
\Rule{\Delta;\Gamma,x:\tau\sj e:\tau'}
     {\Delta;\Gamma\sj \lam{x}{\tau}{e}:\tau\to\tau'}}\\[\GAP]
\clause{
\Rule{\Delta;\Gamma\sj e_1:\tau_1\to\tau_2\quad\Delta;\Gamma\sj e_2:\tau_1}
     {\Delta;\Gamma\sj e_1~e_2:\tau_2}}\\[\GAP]
\clause{
\Rule{\Delta;\Gamma\sj p:\Forn{\alpha}{\tau}{\tau} \quad
      \Delta\sj\tau'_1\subt\tau_1 \quad
      \dots \quad
      \Delta\sj\tau'_n\subt\tau_n}
     {\Delta;\Gamma\sj p~[\tau'_1,\dots,\tau'_n]:
      \tau\{\tau'_1/\alpha_1,\dots,\tau'_n/\alpha_n\}}}\\[\GAP]
\clause{
\Rule{\Delta;\Gamma,x:\sigma\sj e:\tau\quad\Delta;\Gamma\sj p:\sigma}
     {\Delta;\Gamma\sj\Let{x}{p}{e}:\tau}
}
\end{display}

\begin{display}{Judgment {\mdseries $\Delta;\Gamma\sj p:\sigma$}:}
\clause{
\RuleSide{\Delta\sj\Gamma~\mbox{ctxt} \quad
          \Delta,\alpha_1\subt\tau_1,\dots,\alpha_n\subt\tau_n;\Gamma\sj e:\tau}
         {\Delta;\Gamma\sj\Lamn{\alpha}{\tau}{e}:
          \Forn{\alpha}{\tau}{\tau}}
         {(\alpha_1,\dots,\alpha_n\not\in\Delta)}
}
\end{display}

\subsection{Proofs}

The proof of soundness for $\scalc$ is mostly standard, relying on
preservation and progress lemmas. For completeness, we present all the 
lemmas needed to derive the proof, but leave most of the
straightforward details to the reader.

\begin{lemma}\label{l:smain}
\begin{itemize}

\item[(a)] Monomorphic expression substitution preserves typing:
\begin{itemize}
\item If $\sjudget{\Delta;\Gamma,x:\tau'}{e}{\tau}$ and $\sjudget{\Delta;\Gamma}{e'}{\tau'}$, then
  $\sjudget{\Delta;\Gamma}{e\{e'/x\}}{\tau}$.
\item If $\sjudget{\Delta;\Gamma,x:\tau'}{p}{\sigma}$ and $\sjudget{\Delta;\Gamma}{e'}{\tau'}$, then
  $\sjudget{\Delta;\Gamma}{p\{e'/x\}}{\tau}$.
\end{itemize}

\item[(b)] Polymorphic expression substitution preserves typing:
\begin{itemize}
\item If $\sjudget{\Delta;\Gamma,x:\sigma'}{e}{\tau}$ and $\sjudget{\Delta;\Gamma}{p'}{\sigma'}$, then
  $\sjudget{\Delta;\Gamma}{e\{p'/x\}}{\tau}$.
\item If $\sjudget{\Delta;\Gamma,x:\sigma'}{p}{\sigma}$ and $\sjudget{\Delta;\Gamma}{p'}{\sigma'}$, then
  $\sjudget{\Delta;\Gamma}{p\{p'/x\}}{\sigma}$.
\end{itemize}

\item[(c)] Type subsumption preserves subtyping:
\begin{itemize}
\item If $\Delta,\alpha\subt\tau'\sj\tau_1\subt\tau_2$, and $\tau''\subt\tau'$ then
  $\Delta,\alpha\subt\tau''\sj\tau_1\subt\tau_2$.
\end{itemize}

\item[(d)] Type subsumption preserves typing:
\begin{itemize}
\item If $\sjudget{\Delta,\alpha\subt\tau';\Gamma}{e}{\tau}$ and $\tau''\subt\tau'$, then
  $\sjudget{\Delta,\alpha\subt\tau'';\Gamma}{e}{\tau}$.
\item If $\sjudget{\Delta,\alpha\subt\tau';\Gamma}{p}{\sigma}$ and $\tau''\subt\tau'$, then
  $\sjudget{\Delta,\alpha\subt\tau'';\Gamma}{p}{\sigma}$.
\end{itemize}

\item[(e)] Type substitution preserves subtyping:
\begin{itemize}
\item If $\Delta,\alpha\subt\tau'\sj\tau_1\subt\tau_2$, then
  $\Delta\sj\tau_1\{\tau'/\alpha\}\subt\tau_2\{\tau'/\alpha\}$.
\end{itemize}

\item[(f)] Type substitution preserves typing:
\begin{itemize}
\item If $\sjudget{\Delta,\alpha\subt\tau';\Gamma}{e}{\tau}$ then
  $\sjudget{\Delta;\Gamma\{\tau'/\alpha\}}{e\{\tau'/\alpha\}}{\tau\{\tau'/\alpha\}}$.
\item If $\sjudget{\Delta,\alpha\subt\tau';\Gamma}{p}{\sigma}$ then
  $\sjudget{\Delta;\Gamma\{\tau'/\alpha\}}{p\{\tau'/\alpha\}}{\sigma\{\tau'/\alpha\}}$.
\end{itemize}

\item[(g)] Canonical forms:
\begin{itemize}
\item If $\sjudget{}{v}{t}$, then $v$ has the form $c$ (for $c \in C$).
\item If $\sjudget{}{v}{\tau_a\to\tau_b}$, then either $v$ has the form $f$
  (for $f \in F$) or $v$ has the form $\lam{x}{\tau_a}{e_a}$.
\item If $\sjudget{}{p}{\ForT{\alpha_1\subto\tau_{a,1},\dots,\alpha_n\subto\tau_{a,n}}{\tau_a}}$, then
  $p$ has the form $\LamT{\alpha_1\subto\tau_{a,1},\dots,\alpha_n\subto\tau_{a,n}}{e_a}$
\end{itemize}
\end{itemize}
\end{lemma}

\begin{proof}
\indent (a) Proceed by simultaneous induction on the derivations
$\sjudget{\Delta;\Gamma,x:\tau'}{e}{\tau}$ and
$\sjudget{\Delta;\Gamma,x:\tau'}{p}{\sigma}$.

(b) Proceed by simultaneous induction on the derivations
$\sjudget{\Delta;\Gamma,x:\sigma'}{e}{\tau}$ and
$\sjudget{\Delta;\Gamma,x:\sigma'}{p}{\sigma}$.

(c) Proceed by induction on the derivation $\Delta,\alpha\subt\tau'\sj\tau_1\subt\tau_2$.

(d) Proceed by simultaneous induction on the derivations
$\sjudget{\Delta,\alpha:\tau';\Gamma}{e}{\tau}$ and $\sjudget{\Delta,\alpha:\tau';\Gamma}{p}{\sigma}$.
We give the one interesting case of the induction. In the primitive
operation case, $e = f$ ($f \in F$), $\FV(\tau) =
\langle\alpha_1,\dots,\alpha_n\rangle$, and for all $\tau^*_1\subt\tau_1,\dots,\tau^*_n\subt\tau_n$, we have
$\tau\{\tau^*_1/\alpha_1,\dots,\tau^*_n/\alpha_n\}\in\pi_F(f)$.  If $\alpha \neq \alpha_i$, the result is
immediate.  If $\alpha = \alpha_i$, then
for all $\tau^*_1\subt\tau_1,\dots,\tau^*_i\subt\tau',\dots,\tau^*_n\subt\tau_n$, 
$\tau\{\tau^*_1/\alpha_1,\dots,\tau^*_i/\alpha,\dots,\tau^*_n/\alpha_n\}\in\pi_F(f)$ and $\Delta\sj\tau''\subt\tau'$
implies that for all $\tau^*_1\subt\tau_1,\dots,\tau^*_i\subt\tau'',\dots,\tau^*_n\subt\tau_n$, 
$\tau\{\tau^*_1/\alpha_1,\dots,\tau^*_i/\alpha_i,\dots,\tau^*_n/\alpha_n\}\in\pi_F(f)$. Thus,
$\sjudget{\Delta,\alpha\subt\tau'';\Gamma}{f}{\tau}$.

(e) Proceed by induction on the derivation $\Delta,\alpha\subt\tau'\sj\tau_1\subt\tau_2$.

(f) Proceed by simultaneous induction on the derivations
$\sjudget{\Delta,\alpha\subt\tau';\Gamma}{e}{\tau}$ and
$\sjudget{\Delta,\alpha\subt\tau';\Gamma}{p}{\sigma}$. We give the
interesting cases of the induction. 

In the primitive operation case, $e = f$ ($f \in F$), $\FV(\tau) =
\langle\alpha_1,\dots,\alpha_n\rangle$, and for all $\tau^*_1\subt\tau_1,\dots,\tau^*_n\subt\tau_n$ we have
$\tau\{\tau^*_1/\alpha_1,\dots,\tau^*_n/\alpha_n\}\in\pi_F(f)$.  Note $f\{\tau'/\alpha\} = f$.  If $\alpha \neq \alpha_i$,
the result is immediate.  If $\alpha = \alpha_i$, then
for all $\tau^*_1\subt\tau_1,\dots,\tau^*_i\subt\tau',\dots,\tau^*_n\subt\tau_n$ we have
$\tau\{\tau^*_1/\alpha_1,\dots,\tau^*_i/\alpha,\dots,\tau^*_n/\alpha_n\}\in\pi_F(f)$, which implies that
for all $\tau^*_1\subt\tau_1,\dots,\tau^*_{i-1}\subt\tau_{i-1},\tau^*_{i+1}\subt\tau_{i+1},\dots,\tau^*_n\subt\tau_n$,
we have
$\tau\{\tau'/\alpha\}\{\tau^*_1/\alpha_1,\dots,\tau^*_{i-1}/\alpha_{i-1},\tau^*_{i+1}/\alpha_{i+1},\dots,\tau^*_n/\alpha_n\}\in\pi_F(f)$.
Thus, $\sjudget{\Delta;\Gamma\{\tau'/\alpha\}}{f}{\tau\{\tau'/\alpha\}}$.

In the type abstraction case, $p =
\LamT{\alpha_1\subto\tau_{a,1},\dots,\alpha_n\subto\tau_{a,n}}{e_a}$, $\sigma =
\ForT{\alpha_1\subto\tau_{a,1},\dots,\alpha_n\subto\tau_{a,n}}{\tau_a}$, and
$\sjudget{\Delta,\alpha\subt\tau',\alpha_1\subt\tau_{a,1},\dots,\alpha_n\subt\tau_{a,n};\Gamma}{e_a}{\tau_a}$.
Assume $\alpha_1,\dots,\alpha_n \neq \alpha$.  Note
$(\LamT{\alpha_1\subto\tau_{a,1},\dots,\alpha_n\subto\tau_{a,n}}{e_a})\{\tau'/\alpha\} =
\LamT{\alpha_1\subto\tau_{a,1},\dots,\alpha_n\subto\tau_{a,n}}{e_a\{\tau'/\alpha\}}$ and
$(\ForT{\alpha_1\subto\tau_{a,1},\dots,\alpha_n\subto\tau_{a,n}}{\tau_a}\{\tau'/\alpha\} =
\ForT{\alpha_1\subto\tau_{a,1},\dots,\alpha_n\subto\tau_{a,n}}{\tau_a\{\tau'/\alpha\}}$ (because type variables
are precluded from the types of quantified type variables).
Furthermore,
$\sjudget{\Delta,\alpha_1\subt\tau_{a,1},\dots,\alpha_n\subt\tau_{a,n},\alpha\subt\tau';\Gamma}{e_a}{\tau_a}$.
By the induction hypothesis,
$\sjudget{\Delta,\alpha_1\subt\tau_{a,1},\dots,\alpha_n\subt\tau_{a,n};\Gamma\{\tau'/\alpha\}}{e_a\{\tau'/\alpha\}}{\tau_a\{\tau'/\alpha\}}$.
Hence,
$\sjudget{\Delta;\Gamma\{\tau'/\alpha\}}{(\LamT{\alpha_1\subto\tau_{a,1},\dots,\alpha_n\subto\tau_{a,n}}{e_a})\{\tau'/\alpha\}}{\sigma\{\tau'/\alpha\}}$.

(g) For the first part, proceed by case analysis of the derivation
$\sjudget{}{v}{t}$. 
For the second part, proceed by case analysis of the derivation
$\sjudget{}{v}{\tau_a\to\tau_b}$. 
For the third part, proceed by case analysis of $p$. 
\end{proof}

\begin{oldtheorem}{t:ssound}
\begin{theorem}
If $\pi_F$ is sound with respect to $\delta$, $\sjudget{}{e}{\tau}$ and $e
\slra^* e'$, then $\sjudget{}{e'}{\tau}$ and either $e'$ is a value or
there exists $e''$ such that $e' \slra e''$.  
\end{theorem}
\end{oldtheorem}
\begin{proof}
This is a standard proof of soundness, relying on progress and
preservation lemmas:
\begin{itemize}

\item Progress: if $\pi_F$ is sound with respect to $\delta$ and
$\sjudget{}{e}{\tau}$, then either $e$ is a value or there exists $e'$
such that $e \slra e'$. This follows by induction on the derivation
$\sjudget{}{e}{\tau}$. The only interesting case is the application
case $e = e_1~e_2$ when $e_1$ has the form $f$ (for $f\in F$) and
$e_2$ is a value. Then $\tau_a\to\tau = t_a \to t$ for $t_a,t \in T$
and $t_a \to t \in \pi_F(f)$ by the typing judgment for primitive
operations. By part 1 of Lemma~\ref{l:smain}(g), $e_2$ has the form
$c$ (for $c \in C$). Hence, $\sjudget{}{f~c}{\tau}$ and $\delta(f,c)$
is defined by the definition of $\pi_F$ sound with respect to
$\delta$, and the primitive step applies to $e$. 

\item Preservation: if $\pi_F$ is sound with respect to $\delta$, 
$\sjudget{}{e}{\tau}$ and $e \slra e'$, then $\sjudget{}{e'}{\tau}$. 
This follows by induction on the derivation $\sjudget{}{e}{\tau}$.
The only interesting case is the application case 
$e = e_1~e_2$, when $e_1 = f$ (for $f \in F$), $e_2
= c$ (for $c \in C$), and $e' = \delta(f,c)$. the result follows by
the definition of $\pi_F$ sound with respect to $\delta$.

\end{itemize}

To prove soundness, we assume $\sjudget{}{e}{\tau}$ and $e \slra^*
e'$.  Then $e \slra^n e'$ for some $n$.  Proceed by induction on $n$.
In the base case, the theorem is equivalent to progress.  In the step case,
the inductive hypothesis, preservation, and progress suffice to prove
the theorem. 
\end{proof}

\section{The Calculus $\tcalc$}
\label{a:tcalc}

\begin{display}{Value:}
\category{v}{values}\\
\entry{c}{base constant $(c\in C)$}\\
\entry{f}{primitive operation $(f\in F)$}\\
\entry{\lam{x}{\tau}{e}}{functional abstraction}
\end{display}

\begin{display}{Evaluation Contexts:}
\category{E}{evaluation contexts}\\
\entry{[~]}{empty context}\\
\entry{E~e}{application context}\\
\entry{v~E}{argument context}\\
\entry{E~[\tau_1,\dots,\tau_n]}{type application context}\\
\entry{\Let{x}{E}{e}}{local binding context}
\end{display}

\begin{display}{Operational Semantics:}
\clause{(\lam{x}{\tau}{e})~v \tlra e\{v/x\}}\\
\clause{(\LamT{\alpha_1,\dots,\alpha_n}{e})~[\tau'_1,\dots,\tau'_n] \tlra e\{\tau'_1/\alpha_1,\dots,\tau'_n/\alpha_n\}}\\
\clause{\Let{x}{p}{e} \tlra  e\{p/x\}}\\
\clause{f~c \tlra  c' \mbox{~~iff~~} \delta(f,c)=c'}\\
\clause{E[e_1]\tlra E[e_2] \mbox{~~iff~~} e_1\tlra e_2}
\end{display}
The function $\delta: F \times C \rightharpoonup C$ is a partial function defining the
result of applying a primitive operation to a base constant.

\begin{display}{Typing Contexts:}
\category{\Gamma}{type environments}\\
\entry{\cdot}{empty}\\
\entry{\Gamma,x:\tau}{type}\\
\entry{\Gamma,x:\sigma}{type scheme}\\
\category{\Delta}{type variable environments}\\
\entry{\cdot}{empty}\\
\entry{\Delta,\alpha}{type variable}
\end{display}

\begin{display}{Judgments:}
\clause{\Delta\tj \Gamma~\mbox{ctxt}}{good context $\Gamma$}\\
\clause{\Delta\tj \tau~\mbox{type}}{good type $\tau$}\\
\clause{\Delta\tj \sigma~\mbox{scheme}}{good type scheme $\sigma$}\\
\clause{\Delta;\Gamma\tj e:\tau}{good expression $e$ with type $\tau$}\\
\clause{\Delta;\Gamma\tj p:\sigma}{good expression $p$ with type scheme $\sigma$}
\end{display}

\begin{display}{Judgment {\mdseries $\Delta\tj \Gamma~\mbox{ctxt}$}:}
\clause{
\Rule{}{\Delta\tj\cdot~\mbox{ctxt}}
\quad\Rule{\Delta\tj\Gamma~\mbox{ctxt}}
       {\Delta\tj\Gamma,x:\tau~\mbox{ctxt}}
\quad\Rule{\Delta\tj\Gamma~\mbox{ctxt}}
       {\Delta\tj\Gamma,x:\sigma~\mbox{ctxt}}}
\end{display}

\begin{display}{Judgment {\mdseries $\Delta\tj \tau~\mbox{type}$}:}
\clause{
\Rule{}{\Delta \tj t~\mbox{type}} \quad
\Rule{\tj \Delta~\mbox{ctxt} \quad
      \alpha \in \mbox{dom}(\Delta)}
     {\Delta \tj \alpha~\mbox{type}} \quad
\Rule{\Delta \tj \tau_1~\mbox{type} \quad
      \Delta \tj \tau_2~\mbox{type}}
     {\Delta \tj \tau_1\to\tau_2~\mbox{type}}}
\end{display}

\begin{display}{Judgment {\mdseries $\Delta\tj \sigma~\mbox{scheme}$}:}
\clause{
\Rule{\Delta,\alpha_1,\dots,\alpha_n \tj \tau~\mbox{type}}
     {\Delta \tj \ForT{\alpha_1,\dots,\alpha_n}{\tau}~\mbox{scheme}}}
\end{display}

\begin{display}{Judgment {\mdseries $\Delta;\Gamma\tj e:\tau$}:}
\clause{
\RuleSide{}{\Delta;\Gamma\tj c:\pi_C(c)}{(c\in C)}}\\[\GAP]
\clause{
\RuleSide{\text{For all $\tau'\in\pi_C(C)$ such that}\\
          \mathit{unify}(\tau_1,\tau')=\langle(\alpha_1,\tau'_1),\dots,(\alpha_n,\tau'_n),\dots\rangle,\\
          (\tau_1\to\tau_2)\{\tau'_1/\alpha_1,\dots,\tau'_n/\alpha_n\} \in \pi_F(f)}
         {\Delta,\alpha_1,\dots,\alpha_n;\Gamma\tj f:\tau_1\to\tau_2}
         {\left(\begin{array}{c}
                f\in F,\\
                \FV(\tau_1)=\langle\alpha_1,\dots,\alpha_n\rangle
                \end{array}\right)}
}\\[\GAP]
\clause{
\Rule{\Delta\tj\Gamma~\mbox{ctxt}}{\Delta;\Gamma,x:\tau\tj x:\tau} \quad
\Rule{\Delta;\Gamma,x:\tau\tj e:\tau'}
     {\Delta;\Gamma\tj \lam{x}{\tau}{e}:\tau\to\tau'}}\\[\GAP]
\clause{
\Rule{\Delta;\Gamma\tj e_1:\tau_1\to\tau_2\quad\Delta;\Gamma\tj e_2:\tau_1}
     {\Delta;\Gamma\tj e_1~e_2:\tau_2}}\\[\GAP]
\clause{
\Rule{\Delta;\Gamma\tj p:\ForT{\alpha_1,\dots,\alpha_n}{\tau}}
     {\Delta;\Gamma\tj p~[\tau_1,\dots,\tau_n]:
      \tau\{\tau_1/\alpha_1,\dots,\tau_n/\alpha_n\}} \quad
\Rule{\Delta;\Gamma,x:\sigma\tj e:\tau\quad\Delta;\Gamma\tj p:\sigma}
     {\Delta;\Gamma\tj\Let{x}{p}{e}:\tau}}
\end{display}

\begin{display}{Judgment {\mdseries $\Delta;\Gamma\tj p:\sigma$}:}
\clause{
\RuleSide{\Delta\tj\Gamma~\mbox{ctxt} \quad
          \Delta,\alpha_1,\dots,\alpha_n;\Gamma\tj e:\tau}
         {\Delta;\Gamma\tj\LamT{\alpha_1,\dots,\alpha_n}{e}:
          \ForT{\alpha_1,\dots,\alpha_n}{\tau}}
         {(\alpha_1,\dots,\alpha_n\not\in\Delta)}}
\end{display}

\subsection{Proofs}

The proof of soundness for $\tcalc$ is mostly standard, relying on
preservation and progress lemmas. As we did for $\scalc$,  we present
all the  lemmas needed to derive the proof, but leave most of the 
straightforward details to the reader.

\begin{lemma}\label{l:tmain}
\begin{itemize}
\item[(a)] Monomorphic expression substitution preserves typing:
\begin{itemize}
\item If $\tjudget{\Delta;\Gamma,x:\tau'}{e}{\tau}$ and $\tjudget{\Delta;\Gamma}{e'}{\tau'}$, then
  $\tjudget{\Delta;\Gamma}{e\{e'/x\}}{\tau}$.
\item If $\tjudget{\Delta;\Gamma,x:\tau'}{p}{\sigma}$ and $\tjudget{\Delta;\Gamma}{e'}{\tau'}$, then
  $\tjudget{\Delta;\Gamma}{p\{e'/x\}}{\tau}$.
\end{itemize}

\item[(b)] Polymorphic expression substitution preserves typing:
\begin{itemize}
\item If $\tjudget{\Delta;\Gamma,x:\sigma'}{e}{\tau}$ and $\tjudget{\Delta;\Gamma}{p'}{\sigma'}$, then
  $\tjudget{\Delta;\Gamma}{e\{p'/x\}}{\tau}$.
\item If $\tjudget{\Delta;\Gamma,x:\sigma'}{p}{\sigma}$ and $\tjudget{\Delta;\Gamma}{p'}{\sigma'}$, then
  $\tjudget{\Delta;\Gamma}{p\{p'/x\}}{\sigma}$.
\end{itemize}

\item[(c)] Type substitution preserves typing:
\begin{itemize}
\item If $\tjudget{\Delta,\alpha;\Gamma}{e}{\tau}$ then
  $\tjudget{\Delta;\Gamma\{\tau'/\alpha\}}{e\{\tau'/\alpha\}}{\tau\{\tau'/\alpha\}}$.
\item If $\tjudget{\Delta,\alpha;\Gamma}{p}{\sigma}$ then
  $\tjudget{\Delta;\Gamma\{\tau'/\alpha\}}{p\{\tau'/\alpha\}}{\sigma\{\tau'/\alpha\}}$.
\end{itemize}

\item[(d)] Canonical forms:
\begin{itemize}
\item If $\tjudget{}{v}{\mathsf{T}~\tau}$, then $v$ has the form $c$
(for $c \in C$).
\item If $\tjudget{}{v}{\tau_a\to\tau_b}$, then either $v$ has the form $f$
  (for $f \in F$) or $v$ has the form $\lam{x}{\tau_a}{e_a}$.
\item If $\tjudget{}{p}{\ForT{\alpha_1,\dots,\alpha_n}{\tau_a}}$, then $p$ has the form
  $\LamT{\alpha_1,\dots,\alpha_n}{e_a}$
\end{itemize}
\end{itemize}
\end{lemma}

\begin{proof}
(a) Proceed by simultaneous induction on the derivations
$\tjudget{\Delta;\Gamma,x:\tau'}{e}{\tau}$ and
$\tjudget{\Delta;\Gamma,x:\tau'}{p}{\sigma}$.

(b) Proceed by simultaneous induction on the derivations
$\tjudget{\Delta;\Gamma,x:\sigma'}{e}{\tau}$ and
$\tjudget{\Delta;\Gamma,x:\sigma'}{p}{\sigma}$. 

(c) Proceed by simultaneous induction on the derivations
$\tjudget{\Delta,\alpha;\Gamma}{e}{\tau}$ and
$\tjudget{\Delta,\alpha;\Gamma}{p}{\sigma}$. We prove the interesting
cases of the induction. 

In the primitive
operation case, $e = f$ ($f \in F$), $\tau=\tau_1\to\tau_2$,
$\FV(\tau_1) = \langle\alpha_1,\dots,\alpha_n\rangle$, and for all
$\tau_*\in\pi_C(C)$ such that
$\mathit{unify}(\tau_*,\tau_1)=\langle(\alpha_1,\tau'_1),\dots,(\alpha_n,\tau'_n),\dots\rangle$,
we have
$\tau_1\to\tau_2\{\tau'_1/\alpha_1,\dots,\tau'_n/\alpha_n\}\in\pi_F(f)$.
Note $f\{\tau'/\alpha\} = f$.  If $\alpha \neq
\alpha_i$ for any $i$, the result is immediate.  If $\alpha = \alpha_i$ for some $i$
(without loss of generality, let $i=1$), then for any
$\tau_*\in\pi_C(C)$ such that
$\mathit{unify}(\tau_*,\tau_1\{\tau'/\alpha_1\}) =
\langle(\alpha_2,\tau'_2),\dots,(\alpha_n,\tau'_n),\dots\rangle$, we have $\mathit{unify}(\tau_*,\tau_1) =
\langle(\alpha_1,\tau'),(\alpha_2,\tau'_2),\dots,(\alpha_n,\tau'_n),\dots\rangle$, so that
$(\tau_1\to\tau_2)\{\tau'/\alpha\}\{\tau'_2/\alpha_2,\dots,\tau'_n/\alpha_n\}
=(\tau_1\to\tau_2)\{\tau'/\alpha_1,\tau'_2/\alpha_2,\dots,\tau'_n/\alpha_n\}\in\pi_F(f)$.
Thus, $\tjudget{\Delta;\Gamma\{\tau'/\alpha\}}{f}{\tau\{\tau'/\alpha\}}$.

In the type abstraction case, $p = \LamT{\alpha_1,\dots,\alpha_n}{e_a}$, $\sigma =
\ForT{\alpha_1,\dots,\alpha_n}{\tau_a}$, and $\tjudget{\Delta,\alpha,\alpha_1,\dots,\alpha_n;\Gamma}{e_a}{\tau_a}$.
Assume $\alpha_1,\dots,\alpha_n \neq \alpha$.  Note $(\LamT{\alpha_1,\dots,\alpha_n}{e_a})\{\tau'/\alpha\} =
\LamT{\alpha_1,\dots,\alpha_n}{e_a\{\tau'/\alpha\}}$ and $(\ForT{\alpha_1,\dots,\alpha_n}{\tau_a})\{\tau'/\alpha\} =
\ForT{\alpha_1,\dots,\alpha_n}{\tau_a\{\tau'/\alpha\}}$ (because type variables are precluded from the
types of quantified type variables).  Furthermore,
$\tjudget{\Delta,\alpha_1,\dots,\alpha_n,\alpha;\Gamma}{e_a}{\tau_a}$.  By the induction
hypothesis, $\tjudget{\Delta,\alpha_1,\dots,\alpha_n;\Gamma\{\tau'/\alpha\}}{e_a\{\tau'/\alpha\}}{\tau_a\{\tau'/\alpha\}}$.
Hence,
$\tjudget{\Delta;\Gamma\{\tau'/\alpha\}}{(\LamT{\alpha_1,\dots,\alpha_n}{e_a})\{\tau'/\alpha\}}{\sigma\{\tau'/\alpha\}}$.

(d) For the first part, proceed by case analysis of the derivation
$\tjudget{}{v}{\mathsf{T}~\tau}$. 
For the second part, proceed by case analysis of the derivation
$\tjudget{}{v}{\tau_a\to\tau_b}$. 
For the third part, proceed by case analysis of $p$. 
\end{proof}

\begin{oldtheorem}{t:tsound}
\begin{theorem}
If $\pi_F$ is sound with respect to $\delta$, $\tjudget{}{e}{\tau}$ and $e
\tlra^* e'$, then $\tjudget{}{e'}{\tau}$ and either $e'$ is a value or
there exists $e''$ such that $e' \tlra e''$.
\end{theorem}
\end{oldtheorem}
\begin{proof}
This is a standard proof of soundness, relying on progress and
preservation lemmas:
\begin{itemize}

\item Progress: if $\pi_F$ is sound with respect to $\delta$ and
$\tjudget{}{e}{\tau}$, then either $e$ is a value or there exists $e'$
such that $e \tlra e'$. This follows by induction on the derivation
$\tjudget{}{e}{\tau}$. The interesting case of the induction is the
application case $e = e_1~e_2$ when $e_1$ has the form $f$ (for $f \in
F$) and $e_2$ is a value. Then, $\tau_a\to\tau = t_a \to t$ for $t_a,t
\in T$ and $t_a \to t \in \pi_F(f)$ by the typing judgment for
primitive operations.  By part 1 of Lemma~\ref{l:tmain}, $e_2$
has the form $c$ (for $c \in C$).  Hence, $\tjudget{}{f~c}{\tau}$ and 
$\delta(f,c)$ is defined by the definition of $\pi_F$ sound with respect
to $\delta$.  Thus, the primitive step applies to $e$.

\item Preservation: if $\pi_F$ is sound with respect to $\delta$,
$\tjudget{}{e}{\tau}$ and $e \tlra e'$, then $\tjudget{}{e'}{\tau}$. 
This follows by induction on the derivation $\tjudget{}{e}{\tau}$. 
Again, the interesting case is the application case $e = e_1~e_2$ when
$e_1 = f$ (for $f \in F$), $e_2 = c$ (for $c \in C$), and $e' =
\delta(f,c)$.  The result follows by the definition of $\pi_F$ sound
with respect to $\delta$.
\end{itemize}

To prove soundness, we assume $\tjudget{}{e}{\tau}$ and $e \tlra^*
e'$.  Then $e \tlra^n e'$ for some $n$.  Proceed by induction on $n$.
In the base case, the theorem is equivalent to
progress.  In the step case, 
the inductive hypothesis, preservation, and progress suffice to prove
the theorem. 
\end{proof}

\section{Translation Proofs}\label{a:translation}

\begin{oldtheorem}{t:safe} 
\begin{theorem}
If $\pi_F$ is sound with respect to $\delta$ in $\scalc$, then
$\cT\intension{\pi_F}$ is sound with respect to $\delta$ in $\tcalc$.
\end{theorem}
\end{oldtheorem}
\begin{proof}
We need to show that for all $f\in F$ and $c\in C$ such that $\tj
f~c:\tau$ for some $\tau$, then $\delta(f,c)$ is defined, and that
$\cT\intension{\pi_C}(\delta(f,c))=\tau$. Given $f\in F$ and $c\in C$,
assume that $\tj f~c:\tau$. This means that $\tj f:\tau'\to\tau$ and that
$\tj c:\tau'$. From $\tj f:\tau'\to\tau$, we derive that for all
$\tau^*\in\cT\intension{\pi_C}(C)$ such that $\mathit{unify}(\tau^*,\tau')\neq\emptyset$
(since $\tau'$ and $\tau^*$ are both closed types),
$\tau'\to\tau\in\cT\intension{\pi_F}(f)$. By definition of $\cT$, and by
assumption on the form of $\pi_F$, this means that $\tau'\to\tau$ is of the form
$\cT\intension{t'}\to\cT\intension{t}$, with $\cT\intension{t'}=\tau'$ and
$\cT\intension{t}=\tau$. Hence, $\sj f:t'\to t$. From $\tj c:\tau'$, we
derive that
$\cT\intension{t'}=\tau'=\cT\intension{\pi_C}(c)=\cT\intension{\pi_C(c)}$.
Hence, $\pi_C(c)=t'$, and $\sj c:t'$. We can therefore infer that
$\sj f~c:t$. Therefore, by soundness of $\pi_F$ with respect to $\delta$
in $\scalc$, we get that $\delta(f,c)$ is defined, and that
$\pi_C(\delta(f,c))=t$. Thus,
$\cT\intension{\pi_C}(\delta(f,c))=\cT\intension{\pi_C(\delta(f,c))}=\cT\intension{t}=\tau$,
as required.
\end{proof}

The following lemma, relating the correctness of the subtype encoding
and substitution, is used in the proof of Theorem~\ref{t:correct}.
\begin{lemma}\label{l:substitution} 
For all $t$,$t'$, and $\tau$ with $\FV(\tau)\subseteq\langle\alpha\rangle$, if $t^A=\abst{t}$,
$\FV(t^A)=\langle\alpha_1,\dots,\alpha_n\rangle$, and
$\mathit{unify}(\conc{t'},t^A)=\langle(\alpha_1,\tau_1),\dots,(\alpha_n,\tau_n),\dots\rangle$, then
$\cT\intension{\tau\{t'/\alpha\}}=\cT\intension{\tau}[\alpha\mapsto t^A]\{\tau_1/\alpha_1,\dots,\tau_n/\alpha_n\}$.
\end{lemma} 
\begin{proof}
We proceed by induction on the structure of $\tau$.

For $\tau=\alpha$, we immediately get that
$\cT\intension{\tau\{t'/\alpha\}}=\cT\intension{t'}=\conc{t'}$.  Moreover, we
have $\cT\intension{\tau'}[\alpha\mapsto t^A]\{\tau_1/\alpha_1,\dots,\tau_n/\alpha_n\} =
t^A\{\tau_1/\alpha_1,\dots,\tau_n/\alpha_n\} = \conc{t'}$, by the assumption on the
unification of $\conc{t'}$ and $t^A$.

For $\tau=t^*$ for some $t^*$, then
$\cT\intension{\tau\{t'/\alpha\}}=\cT\intension{\tau}=\conc{t^*}$.  Moreover,
$\cT\intension{\tau}[\alpha\mapsto t^A]\{\tau_1/\alpha_1,\dots,\tau_n/\alpha_n\} = \cT\intension{t^*}[\alpha\mapsto
t^A]\{\tau_1/\alpha_1,\dots,\tau_n/\alpha_n\} = \conc{t^*}\{\tau_1/\alpha_1,\dots,\tau_n/\alpha_n\} = \conc{t^*}$.

Finally, for $\tau=\tau'\to\tau''$, we have $\cT\intension{(\tau'\to\tau'')\{t'/\alpha\}} =
\cT\intension{\tau'\{t'/\alpha\}\to\tau''\{t'/\alpha\}} =
\cT\intension{\tau'\{t'/\alpha\}}\to\cT\intension{\tau''\{t'/\alpha\}}$.  By applying the
induction hypothesis, this is equal to $\cT\intension{\tau'}[\alpha\mapsto
t^A]\{\tau_1/\alpha_1,\dots,\tau_n/\alpha_n\}\to\cT\intension{\tau''}[\alpha\mapsto
t^A]\{\tau_1/\alpha_1,\dots,\tau_n/\alpha_n\}=\cT\intension{\tau'\to\tau''}[\alpha\mapsto
t^A]\{\tau_1/\alpha_1,\dots,\tau_n/\alpha_n\}$, as required.
\end{proof}

\begin{oldtheorem}{t:correct} 
\begin{theorem}
If $\sjudget{}{e}{\tau}$, then
$\tjudget{}{\cE\intension{\sjudget{}{e}{\tau}}}{\cT\intension{\tau}}$.
\end{theorem}
\end{oldtheorem}
\begin{proof}
We prove a more general form of this theorem, namely that if $\Delta;\Gamma\sj
e:\tau$, then
$\cT\intension{\Delta};\cT\intension{\Gamma}\rho_\Delta\tj\cE\intension{\Delta;\Gamma\sj
  e:\tau}\rho_\Delta:\cT\intension{\tau}\rho_\Delta$, where:
\[
\cT\intension{\alpha_1\subt\tau_1,\dots,\alpha_n\subt\tau_n} \teq 
\begin{prog}
 \alpha_{11},\dots,\alpha_{1k_1},\dots,\alpha_{n1},\dots,\alpha_{nk_n}\\
 \quad\mbox{where $\tau_i^A=\cA\intension{\tau_i}$}\\
 \quad \mbox{and $\FV(\tau_i^A)=\langle\alpha_{i1},\dots,\alpha_{ik_i}\rangle$}
\end{prog}
\]
and for $\Delta$ of the form $\alpha_1\subt\tau_1,\dots,\alpha_n\subt\tau_n$,
\[
\rho_\Delta \teq \{\alpha_1\mapsto\tau_1^A,\dots,\alpha_n\mapsto\tau_n^A\}.
\]
Similarly, we show that if $\Delta;\Gamma\sj p:\sigma$, then
$\cT\intension{\Delta};\cT\intension{\Gamma}\rho_\Delta\tj\cE\intension{\Delta;\Gamma\sj
  p:\sigma}\rho_\Delta:\cT\intension{\sigma}\rho_\Delta$. We establish this by simultaneous
induction on the derivations $\sjudget{\Delta;\Gamma}{e}{\tau}$ and
$\sjudget{\Delta;\Gamma}{p}{\sigma}$.

For variables, $\Delta,\Gamma\sj x:\tau$ implies that $x:\tau$ is in $\Gamma$. Hence,
$x:\cT\intension{\tau}\rho_\Delta$ is in $\cT\intension{\Gamma}\rho_\Delta$. Hence,
$\cT\intension{\Delta};\cT\intension{\Gamma}\rho_\Delta\tj x:\cT\intension{\tau}\rho_\Delta$.
Similarly for $\Delta;\Gamma\sj x:\sigma$.

For constants $c\in C$, if $\Delta;\Gamma\sj c:\tau$, then we have $\pi_C(c)=\tau$.
Hence, $\cT\intension{\pi_C(c)}\rho_\Delta=\cT\intension{\tau}\rho_\Delta$, and by
definition, $\cT\intension{\pi_C}\rho_\Delta(c)=\cT\intension{\tau}\rho_\Delta$.  This
implies $\cT\intension{\Delta};\cT\intension{\Gamma}\tj c:\cT\intension{\tau}\rho_\Delta$.

For operations $f\in F$, if $\Delta,\alpha_1\subt\tau_1,\dots,\alpha_n\subt\tau_n;\Gamma\sj f:\tau'\to\tau$
(where $\FV(\tau')=\langle\alpha_1,\dots,\alpha_n\rangle$). Hence, for all $\tau'_i\subt \tau_i$, we have
$(\tau'\to\tau)\{\tau'_1/\alpha_1,\dots,\tau'_n/\alpha_n\}\in\pi_F(f)$.  Note that this implies that
each $\tau'_i$ is of the form $t'_i$ for some $t'_i$, due to the
restrictions imposed on $\pi_F$. Furthermore, also due to the
restrictions imposed on $\pi_F$, we must have that $\tau'$ is either $t'$
for some $t'$, or a type variable $\alpha_1$. We need to show that for all
$\tau^*\in\cT\intension{\pi_C}(C)$, if
$\mathit{unify}(\tau^*,\cT\intension{\tau'})=\langle(\alpha_1,\tau_1),\dots,(\alpha_n,\tau_n),\dots\rangle$,
then $\cT\intension{\tau'\to\tau}\{\tau_1/\alpha_1,\dots,\tau_n/\alpha_n\}\in\cT\intension{\pi_F}(f)$.
Take an arbitrary $\tau^*\in\cT\intension{\pi_C}(C)$. By restrictions on
$\pi_C$, $\tau^*$ is of the form $\conc{t^*}$ for some $t^*\in\pi_C(C)$. Now,
consider the different forms of $\tau'$. In the case $\tau'=t'$, we have
$\cT\intension{\tau'}=\conc{t'}$, so that if
$\mathit{unify}(\conc{t^*},\conc{t'})$, then $t^*=t'$. Moreover,
because we assumed that concrete encodings did not introduce free type
variables, then $\FV(\tau')=\emptyset$. Thus,
$\cT\intension{\tau'\to\tau}=\cT\intension{\tau'}\to\cT\intension{\tau}\in\cT\intension{\pi_F}(f)$
follows immediately from the fact that $\tau'\to\tau\in\pi_F(f)$. In the case that
$\tau'=\alpha_1$, then $\cT\intension{\tau'}\rho=\abst{t'_1}$. Let
$\FV(t'_1)=\langle\alpha_{11},\dots,\alpha_{1k_1}\rangle$. Assume
$\mathit{unify}(\conc{t^*},{t'}^A)=\langle(\alpha_{11},\tau_1),\dots,(\alpha_{1k_1},\tau_{k_1}),\dots\rangle$.
Because the encoding is respectful,
$\mathit{unify}(\conc{t^*},{t'}^A)\neq\emptyset$ if and only if $t^*\leq t_1$, that
is, $t^*\subt t_1$. By assumption, we have $(\tau'\to\tau)\{t^*/\alpha_1\}\in\pi_F(f)$.
Therefore, $\cT\intension{(\tau'\to\tau)\{t^*/\alpha_1\}}\in\cT\intension{\pi_F}(f)$.  By
Lemma~\ref{l:substitution},
$\cT\intension{(\tau'\to\tau)\{t^*/\alpha_1\}}=\cT\intension{\tau'\to\tau}\{\tau_1/\alpha_{11},\dots,\tau_{k_1}/\alpha_{1k_1}\}$,
and the result follows. Since $\tau^*$ was arbitrary, we can therefore
infer that
$\cT\intension{\Delta},\alpha_{11},\dots,\alpha_{1k_1},\dots,\alpha_{n1},\dots,\alpha_{nk_n};\cT\intension{\Gamma}\rho_\Delta[\alpha_i\mapsto\tau_i^A]\tj
f : \cT\intension{\tau'\to\tau}\rho_\Delta[\alpha_i\mapsto\tau_i^A]$.

For abstractions, if $\Delta;\Gamma\sj\lam{x}{\tau'}{e}:\tau'\to\tau$, then $\Delta;\Gamma,x:\tau'\sj
e:\tau$. By the induction hypothesis,
$\cT\intension{\Delta};\cT\intension{\Gamma}\rho_{\Delta};x:\cT\intension{\tau'}\rho_\Delta\tj
\cE\intension{e}\rho_\Delta:\cT\intension{\tau}\rho_\Delta$, from which one can infer
that $\cT\intension{\Delta};\cT\intension{\Gamma}\rho_\Delta\tj
\lam{x}{\cT\intension{\tau'}\rho_\Delta}{\cE\intension{e}\rho_\Delta}:\cT\intension{\tau'}\rho_\Delta\to\cT\intension{\tau}\rho_\Delta$,
which yields $\cT\intension{\Delta};\cT\intension{\Gamma}\rho_\Delta\tj
\lam{x}{\cT\intension{\tau'}\rho_\Delta}{\cE\intension{e}\rho_\Delta}:\cT\intension{\tau'\to\tau}\rho_\Delta$.

For applications, if $\Delta;\Gamma\sj e_1~e_2:\tau$, then for some $\tau'$, $\Delta;\Gamma\sj
e_1:\tau'\to\tau$ and $\Delta;\Gamma\sj e_2:\tau'$. By the induction hypothesis, we have
$\cT\intension{\Delta};\cT\intension{\Gamma}\rho_\Delta\tj\cE\intension{e_1}:\cT\intension{\tau'\to\tau}$,
so that
$\cT\intension{\Delta};\cT\intension{\Gamma}\rho_\Delta\tj\cE\intension{e_1}:\cT\intension{\tau'}\to\cT\intension{\tau}$,
and
$\cT\intension{\Delta};\cT\intension{\Gamma}\rho_\Delta\tj\cE\intension{e_2}:\cT\intension{\tau'}$.
This yields that
$\cT\intension{\Delta};\cT\intension{\Gamma}\rho_\Delta\tj(\cE\intension{e_1}\rho_\Delta)~\cE\intension{e_2}\rho_\Delta:\cT\intension{\tau}\rho_\Delta$.

For local bindings, if $\Delta;\Gamma\sj\Let{x}{p}{e}:\tau$, then for some $\sigma$ we
have $\Delta;\Gamma,x:\sigma\sj e:\tau$ and $\Delta;\Gamma\sj p:\sigma$. By the induction hypothesis,
we have $\cT\intension{\Delta};\cT\intension{\Gamma}\rho_\Delta,x:\cT\intension{\sigma}\rho_\Delta\tj
\cE\intension{e}:\cT\intension{\tau}\rho_\Delta$ and
$\cT\intension{\Delta};\cT\intension{\Gamma}\rho_\Delta\tj
\cE\intension{p}\rho_\Delta:\cT\intension{\sigma}\rho_\Delta$. Thus, we have
$\cT\intension{\Delta};\cT\intension{\Gamma}\rho_\Delta\tj
\Let{x}{\cE\intension{p}\rho_\Delta}{\cE\intension{e}\rho_\Delta}:\cT\intension{\tau}\rho_\Delta$.

For type applications, we have $\Delta;\Gamma\sj p~[\tau'_1,\dots,\tau'_n]:\tau$. Then for
all $(\alpha_i,\tau_i)$ in $\cB\intension{p}\Gamma$, we have $\Delta;\Gamma\sj p :
\Forn{\alpha}{\tau}{\tau'}$, $\Delta\sj \tau'_i\subt\tau_i$ for all $i$, and
$\tau=\tau'\{\tau'_1/\alpha_1,\dots,\tau'_n/\alpha_n\}$.  By the induction hypothesis,
$\cT\intension{\Delta};\cT\intension{\Gamma}\rho_\Delta\tj
\cE\intension{p}\rho_\Delta:\cT\intension{\Forn{\alpha}{\tau}{\tau'}}\rho_\Delta$.  Let
$\tau_i^A=\cA\intension{\tau_i}$, and $\FV(\tau_i^A)=\langle\alpha_{i1},\dots,\alpha_{ik_i}\rangle$.
Thus, $\cT\intension{\Delta};\cT\intension{\Gamma}\rho_\Delta\tj\cE\intension{p}\rho_\Delta:\ForT{\alpha_{11},\dots,\alpha_{1k_1},\dots,\alpha_{n1},\dots,\alpha_{nk_n}}{\cT\intension{\tau'}\rho_\Delta[\alpha_i\mapsto\tau_i^A]}$. 
We know that $\Delta\sj \tau'_i\subt \tau_i$ for all $i$, so we have that
$\mathit{unify}(\tau_i^A,\cT\intension{\tau_i}\rho_\Delta)=\langle(\alpha_{i1},\tau_{i1}),\dots,(\alpha_{ik_i},\tau_{ik_i}),\dots\rangle$,
and
$\cT\intension{\Delta};\cT\intension{\Gamma}\rho_\Delta\tj\cE\intension{p}\rho_\Delta~[\tau_{11},\dots,\tau_{1k_1},\dots,\tau_{n1},\dots,\tau_{nk_n}]:\cT\intension{\tau'}\rho_\Delta[\alpha_i\mapsto\tau_i^A]$,
as required.

For type abstractions, we have $\Delta;\Gamma\sj\Lamn{\alpha}{\tau}{e}:\Forn{\alpha}{\tau}{\tau}$.
Thus, we have $\Delta\sj\Gamma~\mbox{ctxt}$, that is, the type variables in $\Gamma$
appear in $\Delta$, and moreover $\Delta,\alpha_1\subt\tau_1,\dots,\alpha_n\subt\tau_n;\Gamma\sj e:\tau$.
Let $\tau_i^A=\cA\intension{\tau_i}$ and $\FV(\tau_i^A)=\langle\alpha_{i1},\dots,\alpha_{ik_i}\rangle$,
for $1\leq i\leq n$. Let $\rho'_\Delta=\rho_\Delta[\alpha_1\mapsto\tau_1^A,\dots,\alpha_n\mapsto\tau_n^A]$.  By the
induction hypothesis, we have
$\cT\intension{\Delta},\alpha_{11},\dots,\alpha_{1k_1},\dots,\alpha_{n1},\dots,\alpha_{nk_n};\cT\intension{\Gamma}\rho'_\Delta\tj\cE\intension{e}\rho'_\Delta:\cT\intension{\tau}\rho'_\Delta$.
>From this we can infer that $\cT\intension{\Delta};\cT\intension{\Gamma}\rho'_\Delta\tj
\LamT{\alpha_{11},\dots,\alpha_{1k_1},\dots,\alpha_{n1},\dots,\alpha_{nk_n}}{\cE\intension{e}\rho'_\Delta}:\ForT{\alpha_{11},\dots,\alpha_{1k_1},\dots,\alpha_{n1},\dots,\alpha_{nk_n}}{\cT\intension{\tau}\rho'_\Delta}$,
which is easily seen equivalent to
$\cT\intension{\Delta};\cT\intension{\Gamma}\rho_\Delta\tj
\cE\intension{\Lamn{\alpha}{\tau}{e}}\rho_\Delta:\cT\intension{\Forn{\alpha}{\tau}{\tau}}\rho_\Delta$.
(We can replace $\cT\intension{\Gamma}\rho'_\Delta$ by $\cT\intension{\Gamma}\rho_\Delta$ by the
assumption that $\Gamma$ is a good context in $\Delta$.)
\end{proof}


\begin{thebibliography}{}

\bibitem[\protect\citename{Blume, }2001]{Blume01}
Blume, M. (2001).
\newblock {No-Longer-Foreign}: Teaching an {ML} compiler to speak {C}
  ``natively''.
\newblock  {\em Proceedings of BABEL'01}.
\newblock Electronic Notes in Theoretical Computer Science, vol. 59.1.
\newblock Elsevier Science Publishers.

\bibitem[\protect\citename{Burton, }1990]{Burton90}
Burton, {F. W.} (1990).
\newblock Type extension through polymorphism.
\newblock {\em ACM Transactions on Programming Languages and Systems}, {\bf
  12}(1), 135--138.

\bibitem[\protect\citename{Cardelli {\em et~al.}\relax, }1994]{Cardelli94}
Cardelli, L., Martini, S., Mitchell, J.~C., \& Scedrov, A. (1994).
\newblock An extension of {System} {F} with subtyping.
\newblock {\em Information and Computation}, {\bf 109}(1/2), 4--56.

\bibitem[\protect\citename{Damas \& Milner, }1982]{Damas82}
Damas, L., \& Milner, R. (1982).
\newblock Principal type-schemes for functional programs.
\newblock {\em Pages  207--212 of:} {\em Conference Record of the Ninth Annual
  ACM Symposium on Principles of Programming Languages}.
\newblock ACM Press.

\bibitem[\protect\citename{Elliott {\em et~al.}\relax, }2000]{Elliott00}
Elliott, C., Finne, S., \& {de Moor}, O. (2000).
\newblock Compiling embedded languages.
\newblock  {\em Workshop on Semantics, Applications, and Implementation of
  Program Generation}.

\bibitem[\protect\citename{Finne {\em et~al.}\relax, }1999]{Finne99}
Finne, S., Leijen, D., Meijer, E., \& {Peyton Jones}, S. (1999).
\newblock Calling hell from heaven and heaven from hell.
\newblock {\em Pages  114--125 of:} {\em Proceedings of the 1999 ACM SIGPLAN
  International Conference on Functional Programming}.
\newblock ACM Press.

\bibitem[\protect\citename{Fluet \& Pucella, }2005]{FluetPucella05}
Fluet, M., \& Pucella, R. (2005).
\newblock Practical datatype specializations with phantom types and recursion schemes.
\newblock {\em Pages  203--228 of:} {\em Proceedings of the ACM SIGPLAN Workshop on ML}.
\newblock Elsevier Science Publishers.

\bibitem[\protect\citename{Leijen \& Meijer, }1999]{Leijen99}
Leijen, D., \& Meijer, E. (1999).
\newblock Domain specific embedded compilers.
\newblock {\em Pages  109--122 of:} {\em Proceedings of the Second Conference
  on Domain-Specific Languages (DSL'99)}.

\bibitem[\protect\citename{Milner, }1978]{Milner78}
Milner, R. (1978).
\newblock A theory of type polymorphism in programming.
\newblock {\em Journal of Computer and Systems Sciences}, {\bf 17}(3),
  348--375.

\bibitem[\protect\citename{Milner {\em et~al.}\relax, }1997]{Milner97}
Milner, R., Tofte, M., Harper, R., \& MacQueen, D. (1997).
\newblock {\em The Definition of {Standard ML} (revised)}.
\newblock Cambridge, Mass.: The MIT Press.

\bibitem[\protect\citename{Pessaux \& Leroy, }1999]{Pessaux99}
Pessaux, F., \& Leroy, X. (1999).
\newblock Type-based analysis of uncaught exceptions.
\newblock {\em Pages  276--290 of:} {\em Conference Record of the Twenty-Sixth
  Annual ACM Symposium on Principles of Programming Languages}.
\newblock ACM Press.

\bibitem[\protect\citename{R\'{e}my, }1989]{Remy89}
R\'{e}my, D. (1989).
\newblock Records and variants as a natural extension of {ML}.
\newblock {\em Pages  77--88 of:} {\em Conference Record of the Sixteenth
  Annual ACM Symposium on Principles of Programming Languages}.
\newblock ACM Press.

\bibitem[\protect\citename{Reppy, }1996]{Reppy96}
Reppy, J.~H. (1996).
\newblock {\em A safe interface to sockets}.
\newblock Technical memorandum. AT\&T Bell Laboratories.

\bibitem[\protect\citename{Wand, }1987]{Wand87}
Wand, M. (1987).
\newblock Complete type inference for simple objects.
\newblock  {\em Proceedings of the 2nd Annual IEEE Symposium on Logic in
  Computer Science}.

\end{thebibliography}
\end{document}